\documentclass[referee]{aa}
\usepackage[usenames]{color}
\usepackage{pdflscape}

\def\l{$\lambda$}
\def\mbh{$M_{\rm BH}$\/}
\def\nh{$n_{\mathrm{H}}$\/}
\def\lledd{$L/L_{\rm Edd}$}
\def\lbol{$L_{\rm bol}$}

\def\nc{$N_{\rm c}$\/}
\def\rfe{$R_{\rm FeII}$}

\def\feiiq{\rm Fe{\sc ii}$\lambda$4570\/}

\def\ltsima{$\; \buildrel < \over \sim \;$}
\def\ltsim{\lower.5ex\hbox{\ltsima}}  
\def\gtsima{$\; \buildrel > \over \sim \;$}

\def\gtsim{\lower.5ex\hbox{\gtsima}}

\def\ha{{\sc H}$\alpha$}

\def\civ{{\sc{Civ}}$\lambda$1549\/}

\def\cmq{cm$^{-2}$\/}
\def\cm3{cm$^{-3}$\/}
\def\hb{{\sc{H}}$\beta$\/}

\def\hbbc{{\sc{H}}$\beta_{\rm BC}$\/}

\def\hbnc{{\sc{H}}$\beta_{\rm NC}$\/}
\def\mgii{{Mg\sc{ii}}$\lambda$2800\/}
\def\mgiibc{{Mg\sc{ii}}$\lambda2800_{\mathrm{BC}}$\,}

\def\ciii{{\sc{Ciii]}}$\lambda$1909\/}
\def\oiiiopt{{\sc{[Oiii]}}$\lambda\lambda$4959,5007\/}
\def\o4363{{\sc{[Oiii]}}$\lambda$4363\/}
\def\oii{{\sc{[Oii]}}$\lambda$3727\/}

\def\fei{{Fe{\sc i}}}

\def\feiiuv{{{\sc{Feii}}}$_{\rm UV}$\/}
\def\feiiopt{{Fe \sc{ii}}$_{\rm opt}$\/}
\def\feii{{Fe\sc{ii}}\/}

\def\fe{{\sc{Fe}}\/}

\def\vr{{$v_{\mathrm r}$}}

\def\fe76087{{\sc [Fe vii]}$\lambda$6087\/}
\def\oiii{{\sc [Oiii]}$\lambda$5007}

\def\kms{km~s$^{-1}$}

\def\ergss{ergs s$^{-1}$\/}

\def\heii{{{\sc H}e{\sc ii}}$\lambda$4686\/}

\def\rb{$r_{\rm BLR}$\/}

\def\ledd{$L_{\rm Edd}$\/}

\defcitealias{marzianietal13}{Paper I}

\begin{document}
\title{Is \mgii\ a Reliable Virial Broadening Estimator for  Quasars?}

%

\author{ Paola Marziani
\inst{1}\fnmsep\inst{2}
\and 
Jack W. Sulentic\inst{2}
\and 
Ilse Plauchu-Frayn\inst{2}\fnmsep\thanks{Now at Instituto de Astronom\'\i a, UNAM, Ensenada Campus, Baja California, M\'exico} 
\and  
Ascensi\'on del Olmo\inst{2} }
\institute{INAF, Osservatorio Astronomico di Padova, Padova, Italy\\
\email{paola.marziani@oapd.inaf.it} 
\and 
Instituto de Astrof\'isica de Andaluc\'ia, CSIC, Granada, Espa\~na
}
\date{}
\authorrunning{P. Marziani et al.}
\titlerunning{\mgii\ line broadening}

\abstract
{Broad \mgii\ and \hb\ lines have emerged as the most reliable virial
estimators of black hole mass in quasars. Which is more reliable? Part of
the challenge centers on comparing \mgii\ and \hb\ line profiles
in order to improve the $\pm$ 1  dex \mbh\ uncertainties inherent in
single-epoch FWHM measures from noisy spectra.}
{Comparison of  \mgii\ and \hb\ profile measures in the same sources
and especially FWHM measures that provide the virial broadening estimator.}
{Identification of 680 bright Sloan Digital Sky Survey  DataRelease 7 quasars with spectra showing
both \mgii\ and H$\beta$ lines, at redshift  $0.4 \le z \le 0.75$. The s/n of these spectra are
high
enough to allow binning in the { ``four-dimensional  (4D) eigenvector 1''} optical plane and
construction of high s/n composite spectra.}
{
We confirm that   \mgii\ shows a profile that is $\approx$ 20\%\
narrower as suggested in some previous studies. FWHM measures for
Population B sources  { (i.e., with FWHM of \hb\ larger than 4000 \kms)} are uncertain because they show complex profiles with at least two
broad-line components involving a  nearly unshifted broad  and redshifted
very-broad component. Only the broad component is likely to be a valid virial
estimator.
If \hb\ and \mgii\ are not corrected for the very broad component  then black hole mass \mbh\ values for 
Population B
sources will be systematically overestimated by up to $\Delta\log$\mbh\ $\approx
0.3-0.4$\ dex.
We suggest a simple correction that can be applied to the majority of
sources.
\mgii\  is the safer virial estimator for Population B sources because the
{ centroid shifts with respect to rest frame 
are lower}  than for \hb. In the broad and very broad component  { profile} interpretation this is a
consequence of the lower very broad to broad component intensity ratio for \mgii. Eigenvector-based studies
show that effective discrimination of black hole mass and Eddington ratio at fixed redshift is not achieved
via luminosity binning but rather by binning in a ``4D eigenvector 1'' context that reflects
different broad line region geometry/kinematics likely driven by Eddington ratio.}{}
\maketitle
\voffset=1cm
\section{Introduction}
\label{intro}

{ Estimations of black hole mass (\mbh) and
Eddington ratio (\lledd) for large and diverse samples of quasars
\citep[][and references therein]{marzianisulentic12,shen13}
are needed for both astrophysical and cosmological studies. Until recently
the majority of virial black hole \mbh\ estimates have not taken profile diversity into
account.  Rather, broad-line FWHM measures have been assumed to be valid
virial estimators in all quasars. The situation becomes even more complicated when
we compare FWHM estimates derived from different lines that often show quite
different profile shapes
\citep[e.g.,][]{willsetal93,brothertonetal94,bartheletal90,baskinlaor05b,zamfiretal10,hoetal12}.
The goal of the present paper is to
systematize differences between  \hb\ and \mgii\ in the process of
comparing the relative merits of FWHM \hb\ and \mgii\ as
virial broadening estimators. This is done within the context of a 4D
Eigenvector 1 parameter formalism.}

\subsection{The 4D eigenvector 1 parameter space}

{ Interpretation of quasar spectra in the context of a 4D parameter space
stemming from eigenvector studies reveals a large diversity of line
profile properties \citep{sulenticetal00a,sulenticetal07,marzianietal03b, marzianietal10}  likely driven by \lledd, with \mbh\ and source
orientation also playing important roles \citep{borosongreen92,marzianietal01,yipetal04}. The 4DE1 (4D Eigenvector 1)
formalism owes its inception to a principal component analysis study of
systematic trends in quasar spectra  \citep{borosongreen92} from the the
Palomar-Green (PG) survey. The first eigenvector  (E1)
included a correlation between \oiiiopt\ and optical \feii\ strength and
an anti-correlation between the latter parameter and FWHM H$\beta$.
Variants of E1 were later found in
several studies 
\citep[e.g.][]{marzianietal96,boroson02,grupe04,kuraszkiewiczetal02,sulenticetal00b,sulenticetal02,yipetal04,kovacevicetal10,kruzceketal11,tangetal12,wangetal06}.  E1 was
eventually expanded into 4DE1 parameter space
\citep{sulenticetal00a,sulenticetal07} with the
addition \civ\ profile shift and soft X-ray photon index as important
parameters that show    some of the strongest intercorrelations. The
higher dimensionality involves parameters that are
observationally  independent and that correspond to different physical
processes (see the reviews by \citealt{sulenticetal00a} and \citealt{marzianisulentic12}
for more
detailed discussions). Quasars occupy a well defined sequence in the  ``optical plane'' of 4DE1 defined by the FWHM of the \hb\  broad component and by the intensity ratio between \feiiq\ } { and \hb\ \citep{borosongreen92,sulenticetal02}. The 4DE1 optical plane therefore offers a tool that allows for the definition and identification of quasar populations or spectral types.   }

\subsection{Two main quasar populations}

{  Exploration of the 4DE1 parameter space gave rise to
the concept of two quasar populations that, if not truly
distinct, are an effective way to distinguish important
spectroscopic differences between quasars. Pop. A sources
show a wide range of broad-line widths (FWHM \hb $\sim$ 600
-- 4000\kms) with a majority of sources (in low redshift
samples) between FWHM \hb $\sim$ 1000 -- 4000 \kms. Few
sources with \rfe $>$ 0.5 show FWHM \hb  $>$ 4000 \kms\
which was partial motivation for the Pop. A-B formalism.
Pop. A involves a largely radio-quiet quasar population as
few radio-loud quasars (especially with double-lobe
Fanaroff-Riley II (FRII) radio morphology) show FWHM
\hb $<$4000 \kms\ \citep{zamfiretal08}. AGN with FWHM \hb
$\leq$ 2000 \kms\ are often referred to as narrow line
Seyfert 1 (NLSy1) sources but there is continuity in source
properties over the full Pop. A range of FWHM \hb.
The  ``Population A'' definition
changes in higher redshift samples because the minimum
observed broad-line FWHM of \hb\ slowly increases with source
luminosity (see Figure 11 in \citealt{marzianietal09}, and
\citealt{dultzinetal11}).  Pop. A \hb\ profiles are best fit
with a single symmetric Lorentz function while Pop. B are
not. Following previous results for \hb\ we fit all Pop. A
\mgii\ profiles with Lorentz functions
\citep{veroncettyetal01,sulenticetal02}.
Pop.  B quasars show an even wider range of FWHM \hb\ than Pop. A sources.
The full
observed range is FWHM \hb\ = 4000 -- 40000 \kms\  \citep{wangetal09},
but   values FWHM \hb$>$16000 \kms\ are
extremely rare with almost all sources in a low redshift sample
between  4000 \kms\ and 12000 \kms.  Most Pop. B sources show \rfe
$<$ 0.5. Pop B is a mixed radio-quiet and radio-loud
population with the majority of radio-loud sources within
population B. Pop. B sources are different from Pop. A
because -- most fundamentally -- they show composite \hb\
profiles. The majority of Pop. B \hb\ profiles can be modelled with an (almost) unshifted broad (BC) and redshifted very broad  component (VBC). 
It is usually assumed that
the unshifted BC is the same as the single
component observed in Pop. A sources while the VBC is
something non-virial and apparently unique to Pop. B sources
(both radio-quiet and radio-loud). Further discussion on
Pop. A and B is provided by \citet{sulenticetal11}.}

\subsection{Black hole mass estimates}

The equation used to estimate \mbh\ can be written, under the assumption
that line broadening in quasars is due to virialized motions in the
emitting gas, as:
\begin{equation}\label{eq:vir} M_{\mathrm BH} = {\cal
F}\frac{r_\mathrm{BLR}(\Delta v)^{2}}{G}
\end{equation} where ${\cal F}$ is factor $\approx 1$ that depends on
geometry of the emitting gas \citep[e.g.,][]{onkenetal04,grahametal11},
\rb\ is an emissivity-weighted radial distance of the
broad line emitting region (BLR), and  $\Delta v$\ is the virial
broadening term,
customarily measured from
the width of a suitable emission line. The most accurate \mbh\ estimates
come from reverberation
mapping measures of \rb \citep[e.g.][]{horneetal04,denneyetal09}  coupled
with \hb\ rms profile widths
(FWHM or velocity dispersion $\sigma$; \citealt{petersonetal04}). We now
have $\approx$60 reverberation derived radii
\citep{bentzetal09a,bentzetal10,bentzetal13}.
The cost in telescope time for such accurate  \rb\ and rms
line width estimates is very high even for the brightest sources. It is
clear that
we must rely  on secondary methods in order to
obtain a large numbers of additional \mbh\ estimates, as needed for the
analysis
of an ever increasing population of catalogued quasars
\citep{schneideretal10,rossetal12}.
These methods involve single-epoch FWHM measures
and \rb\ estimates derived from the apparent correlation between
reverberation derived radii and source
luminosity measures (the so-called ``Kaspi relation;''
\citealt{kaspietal05}).
{ We suggest that line measures for large source samples can be better
utilized if 4D parameter space coordinates for each source are taken into
account. This is especially true if 4DE1 coordinates are driven by source
Eddington ratio.}

Important issues connected with the use of Eq. \ref{eq:vir} involve: 1)
the validity of extrapolating the Kaspi relation beyond the range of
redshift and source luminosity
represented in the reverberation sample and 2)  estimation of the virial
broadening $\Delta v$.  There are two sides to virial broadening estimation: a) the
choice of a suitable emission line (assumed primarily broadened by virial
motions) and b) selection of a suitable parameter (FWHM, $\sigma$) to
define the virial broadening $\Delta v$. Single-epoch FWHM \hb\ measures
have provided \mbh\ estimates for large samples at low and high redshifts.
Measures of the second moment $\sigma$ are less useful in the case of a
broad lines with more than one component. Beyond redshift $z \gtsim$ 0.7 we
lose the most trusted source of single-epoch FWHM measures (i.e. \hb).
We can either follow \hb\ into the infrared and/or adopt FWHM measures of
other broad lines as virial estimators. We have adopted the former
approach taking advantage of Very Large Telescope (VLT) ISAAC 
\citep{moorwoodetal98} spectroscopy to follow \hb\ through the IR windows
and obtain
accurate as possible \mbh\ estimates for $\approx$50 luminous quasars in
the range $z \approx 1.0 - 3.0$
\citep{sulenticetal04,sulenticetal06,marzianietal09}. We are currently
processing a sample of $\approx$25 sources where \hb\ is redshifted into
the $K$\ band window ($z \approx$  3.0 -- 3.7) beyond which \hb\ is
lost (Sulentic et al. 2013 in preparation).

\subsection{Use of \mgii}

The strongest available surrogate broad lines that are redshifted into
the optical over a wide $z$\ range are \civ\ and \mgii.  The region of
\ciii\ is a blend of permitted and semi-forbidden lines whose apparent
intricacy has deterred previous workers although there is potential for
progress if moderate resolution  high s/n spectra are available 
\citep{marzianietal10,marzianietal11,negrete11,negreteetal12}. \civ\ is
not recommended as a virial estimator because FWHM \civ\ does not even
show a correlation with FWHM \hb\
\citep{marzianietal96,sulenticetal07,netzeretal07,shenetal08,richardsetal11}.
\civ\ also shows a blueshift and profile asymmetry in many sources
\citep{gaskell82,brothertonetal96,sulenticetal07,richardsetal11}
challenging
the validity of the virial assumption. Some consistency can  be achieved if
several cautions and  corrections are applied to the \civ\ measures
( \citealt{assefetal11,denney12,greeneetal10}, and especially in the 4DE1 context as done by \citealt{sulenticetal07}) but this is not the
road to \mbh\ estimates
with uncertainties less than 1dex. In the 4DE1 context
\citep{sulenticetal00b}
both \hb\ and \civ\ show strikingly different profile properties for
Pop. A
and B sources.  
Such differences cannot be easily quantified
with low resolution and/or low s/n spectra.

The rest wavelength of \mgii\  makes it a potentially valuable
ground-based virial estimator from $z \approx$0.4 up to $z \approx 2.1$.
Rapid
improvements in IR spectroscopy also makes \mgii\ accessible at higher
redshifts. It has been measured in $K$\ band for the quasar with highest
known redshift $z \approx$ 7.085 redshift \citep{mortlocketal11}. Is
\mgii\ a safe \hb\ surrogate? Mg$^{+}$\l\l 2796,2803 is a low ionization
line (LIL) doublet long thought to be produced in a dense optically-thick
region where at least part of the Balmer lines arise
\citep{grandiphillips79,netzer80}. A possible complication is that \mgii\
is a
resonance line and therefore often affected by broad
\citep{gangulybrotherton08} and narrow absorption \citep{wildetal08}.
Several authors have attempted to
define a  relation between \mgii\ and \hb\ to provide consistent \mbh\
estimates from the two lines
\citep{mclurejarvis02,vestergaardpeterson06,wangetal09,shenliu12,trakhtenbrotnetzer12}.
\mgii\ is  assumed  to be a virial estimator as reliable as  \hb\
because their line widths are correlated. Latest studies however show
considerable scatter (and deviation from parity) in the FWHM \hb\ vs. FWHM
\mgii\  correlation.
There is strong evidence that FWHM \mgii\ is about 20\%\ smaller than FWHM
\hb\ \citep{wangetal09} in many sources.

We report here on a more detailed comparison of \hb\ and \mgii\ as virial
estimators. This work goes beyond several recent works  considering
line shape properties (shifts and asymmetries, presence of a very broad
component) that could weaken or invalidate the virial assumption.  Such a
more detailed study requires separate consideration of sources with FWHM
\hb\
less than or greater than 4000 \kms\ (Population A and B respectively; cf.
Pop. 1 and 2 of \citealt{collinetal06}). Results for Pop. A sources were
presented in a companion paper
\citep[][hereafter \citetalias{marzianietal13}]{marzianietal13}. The
majority ($\approx$ 80 \%) of Pop. A sources show an unshifted \mgii\
profile that is significantly less
broad ($\sim$20\%) than \hb\ making it the virial estimator of choice. 
Pop. A \hb\ profiles are best fit with a single symmetric Lorentz
function. Gaussian fits  to Pop. A profiles will result in a systematic
overestimation of \mbh.
This paper focuses mainly on Pop. B sources
where broader and more complex lines are observed.
A subset of  Pop. A sources show a significant \mgii\ blueshift
\citepalias{marzianietal13} making FWHM \hb\ a safer virial estimator.
These sources  can be described as
the youngest least-evolved quasars with strong ongoing star formation mong all quasars in the present sample  
\citep[e.g.,][]{sulenticetal00a,mathur00,wangetal06,sanietal10}.
Extreme values of \rfe\ appear to be an empirical signature of quasar
youth perhaps reflecting extreme low ionization and  metal enrichment as
well \citepalias[c.f.][]{marzianietal13}.

The advent of the { Sloan Digital Sky Survey} (SDSS) makes possible  accurate  direct comparison  of \hb\ and \mgii\ using the
brightest 600+ quasars where {\em
both} lines appear in the spectra (\S \ref{sample} describes the sample of
SDSS quasars). Individually most of these spectra are
too noisy (typically  s/n $<$ 5  in the continuum near the two lines) to
yield good measures and, therefore, a reliable application of \mbh -- FWHM
-- relations. We want to move beyond single-epoch measures of the past 20
years  involving random samples that yield the same result -- quasar black
hole masses  with mean
\mbh  $\sim$ 8.5$\pm$1.5 \citep[see the paradoxical result of
][]{croom11}. We
think that the best path to accomplish this goal involves generation of
median composite spectra  { in the 4DE1 context} \citep[][\S
\ref{compo}]{sulenticetal02}. The composites can provide very high s/n
line profiles from which more accurate measures can be obtained. Results
about line profiles and trends are given in \S \ref{results}.  Our
findings are discussed in \S\
\ref{discuss} that presents the necessary cautions
and calibration factors that can ultimately yield improved \mbh\ values.
We will only briefly analyze (\S \ref{virial}) a luminosity-dependent
scaling law such as the one recently
presented  in \citet{trakhtenbrotnetzer12} as well as the determination of
${\cal F}$ that is a distinct and  very complex problem \citep[e.g.,
][]{collinetal06,netzermarziani10,grahametal11,parketal12}.

\section{Sample Selection and Spectral Binning \label{sample}}

\subsection{Sample Selection}


Spectral binning offers the  possibility to generate much higher s/n composite spectra using the SDSS database if a binning context can be identified. We searched SDSS-DR7 for sources catalogued as type-1 AGN (quasars)  in the redshift range 0.4 -- 0.75 and with magnitudes brighter than $ \approx$18.5 in the $g$, $r$ or $i$\ bands. Below the assumed magnitude 
limit the low s/n of individual spectra make it difficult to estimate even FWHM \hb\ and \rfe\ well enough to permit reliable bin assignments. We also searched the list of \citet{zhouetal06} in order to include type-1 AGN showing broad line FWHM \hb$<$ 1000\kms. Such sources are not identified as quasars in SDSS but show strong FeII emission characteristic of type 1 AGN. The resultant sample consisted of 716 quasars.  The next step involved discarding very noisy spectra and some sources with unusually red colors (obviously reddened quasars where \mgii\ was very weak or absent) thus reducing the sample (the same  one considered in \citetalias{marzianietal13}) to 680 quasars.  Narrow and broad absorption line sources were excluded from the sample unless, in the case of narrow absorption lines, the absorption was judged insufficient to affect our ability to measure broad line binning parameters. 

\subsection{Spectral binning}

Spectral binning  in the 4DE1 optical plane maps sources using measures of FWHM \hb\ vs. \rfe\ = W(\feiiq) / W (\hb) $\approx$ I(\feiiq)/I(\hb), where \hb\ refers to measures of the broad line profile of \hb\ \ (\hbbc) corrected for narrow line \hb\ emission as well as any contamination from \oiiiopt, \heii,  and \feii\ emission \citep{sulenticetal02}. { Recent works, including the present one consider the intensity ratio rather than the ratio of equivalent widths.} Spectral types are defined by binning in intervals of \rfe\ ($\Delta$\rfe = 0.5, from A1 [\rfe$<$0.5]  to A4 [1.5$\le$\rfe$<$2.0]) and FWHM  ($\Delta$FWHM \hb = 4000 \kms, from A1 (FWHM \hb $\le$ 4000 \kms) and B1 to B1$^{++}$ [12000 \kms $\le$ FWHM \hb $<$ 16000 \kms]). Separation  of the sample into spectral types will allow us to avoid the danger of mixing together sources that show very different spectral line profile properties. Insofar as \hb\ and \feii\ measures reflect physical conditions in the BLR, the bins isolate sources with similar values of density, ionization parameter, and \lledd\  (see e.g., \citealt{marzianietal01,marzianietal10}). Binning allows us to also take into account the complexity of line profiles and to separate, at least in an heuristic way, major flux contributions that are due to gas that is not likely to be virialized (i.e., the redshifted very broad component of \hb\ in Population B sources). 
Bin assignments were  made as described in \citetalias{marzianietal13}.  We made a final inspection supplemented by {\sc iraf splot} measures, for sources falling near bin boundaries.  A few dozen sources were moved  into adjacent bins but none of this fine tuning is likely to have  had much effect on resultant median composites. Two bins (A2 and B1) are most densely populated while our binning contains enough sources to permit generation of composites for 8 bins that span the FWHM \hb\ and \rfe\ ranges occupied by the majority of low redshift type 1 sources ({ with an inferred 2 dex range in Edditington ratio, $0.01 \la  \log$ \lledd   $\la $1 that is the one of luminous type-1 AGN;} \citealt{woourry02a,marzianietal03b,kollmeieretal06,davislaor11}). 

Table \ref{tab:numbers} gives the number of sources ($N_\mathrm{tot}$) in each bin. Boldface numbers indicate spectral types for which a composite spectrum was constructed.  
The separation into Pop. A and B sources divides almost evenly   low redshift $z < $1.0 quasar samples (45\% Pop. B. in \citet{zamfiretal10} and 54\% in this study).  Bin B2 was almost devoid of sources in previous studies of low-luminosity sources \citep{marzianietal03a,zamfiretal10}  while it accounts for 3\%\ of the present sample. Sources with B2  spectra are more numerous at high luminosity where they account for $\la$20\%\ of our Very Large Telescope ISAAC  sample \citep{marzianietal09}.  Given the shape of the source distribution in the 4DE1 optical plane and the crudeness of the binning it is likely that bin A1 is a blend of  A2 and B1 sources. The B2 composite is mainly due to spectra of sources whose location in the 4DE1 optical plane is close to the boundary of A2 and B1. In other words it could be that B2 sources  are a physically diverse mixture \citep{marzianietal01}.  More than anything else, B2 provides a hints of the confusing results that one would obtain mixing B1 and A2 spectra. Data and measures on B2 are reported but  scantily discussed. 

\subsection{Radio-Loud Sub-Samples}

The entire sample of 680 binned quasars was cross-correlated against the
{ Faint Images of the Radio Sky at Twenty-Centimeters} (FIRST) catalogue \citep{beckeretal95}. 
Radio maps and  catalogue data
for 163 FIRST-detected sources were used to isolate radio-loud (RL) quasars
from the general radio-quiet (RQ) population. The majority of RL sources
belong to Pop. B \citep{sulenticetal03,zamfiretal08}. Most radio-detected RQ sources
show core-dominated (CD), sometimes core-jet morphology. Sources were classified as RL lobe-dominated (LD) when a visual inspection revealed a complex, usually double-lobe structure.  All sources { except one\footnote{ SDSS J093704.05+293704.9 has   $\log P_{\nu} \approx 31.2$\ at 1.4 GHz but classical FRII morphology.}  have radio power  
$\log P_{\nu} \ge 31.6$ at 1.4 GHz ($P_{\nu}$\ in units of erg s$^{-1}$ Hz$^{-1}$), the nominal lower specific power for Fanaroff-Riley II (FRII) sources. }The radio morphology defines the classical RL phenomenon with the weakest FRII sources providing an empirical lower limit for this activity. Compact sources were considered core-dominated (CD) RL if their rest frame
specific emission   was above the FRII lower limit, as in
\citet{zamfiretal08}.
This corresponds roughly to $R_\mathrm{K}$ =70 as defined in
\citet{kellermannetal89}.
The lower limit of radio power in FRII sources excludes all or most
radio-detected
RQ sources { that appear as CD in  FIRST}. Table \ref{tab:numbers}
reports the number of sources in each bin along with the numbers of FIRST
detections
(N$_\mathrm{FIRST}$). FIRST non-detections (i.e. undetected RQ sources) and
detections are tabulated in columns 3 and 4. FIRST should detect all RL
sources in
the redshift range under consideration. Radio detections were assigned an
FRII or CD morphology  (numbers of sources are in columns 5 and 6) based on inspection of FIRST images.

\section{Generation and Analysis of Composite Spectra \label{compo}}

{ Three steps} preceded generation of composite spectra: { 1) assignment of
sources to spectral bins as described in the previous section,} 2) deredshifting of
source spectra using SDSS $z$\ values, and 3) application of a second-order correction connected with a known bias in SDSS redshift estimates
\citep{hewettwild10}. Accurate rest frame definition can be a serious problem. Knowledge of the rest frame is important for identification and physical interpretation of internal broad and narrow line shifts that show a wide velocity range. They  differ widely  from source to source and depend on ionization state \citep{gaskell82,tytlerfan92}. The technique used for redshift computation in
SDSS is reliable but not accurate enough to yield uncertainties as low
as $\approx 30$ \kms\ \citep{hewettwild10}. The \oiiiopt\ line strength is a
known E1 correlate with \oiiiopt\ becoming weaker, broader, and hence
more difficult to measure as one proceeds from B to the A1 bin and 
along the sequence from bin A1 to A4 \citep{sulenticetal02,zamfiretal10}. Bin A3 and A4 sources show a systematic \oiiiopt\ blueshift with respect to low ionization narrow lines. The shift amplitude in these ``blue outliers'' is also believed to be related to the E1 sequence \citep{zamanovetal02,marzianietal03b,komossaetal08,huetal08a}.

Wavelengths of the three most prominent narrow lines \oii, \hb, and \oiii\
were measured whenever detected. Systematic shifts were computed taking an average of the three lines in each source spectrum and clipping individual
measurements in cases of disagreement because of poor data or intrinsic blueshift of \oiii. This means that in several cases only two lines could be used for redshift determination. In a few cases (especially in  A3 and A4 bins,
where oxygen lines can be very weak) only \hbnc\ could be used or, failing
that, the adopted SDSS $z$\ value. The ``bias'' in the SDSS redshift
determination is larger for extreme A bins but relatively small for Pop. B and bin A1
sources. This is not surprising, since bin B1 shows  stronger and sharper
narrow lines that permit more accurate redshift determinations with the SDSS
algorithm.

After sample selection, bin assignments, and redshift correction we
generated  median composite spectra for \hb\ and \mgii. RL and RQ composites was also constructed for bin B1.
Wavelength uncertainties of composite spectra  were quantified by
measuring a posteriori additional low-ionization narrow lines. Systematic effects were found to be small, { with a radial velocity difference} $\Delta$\vr$\la$ 25
\kms. The dispersion { in radial velocity} is $\sigma_{\Delta{v}_\mathrm{r}}
\la$ 50 \kms\ for all bins in line with estimations of \citet{hewettwild10}. In practice this means that any line is assumed to be unshifted if  $\sigma_{\Delta{v}_\mathrm{r}}
\la$ 100 \kms.

\subsection{\hb\ and \mgii\ spectral range}
\label{hbfit}

Interpretation of the \hb\ and \mgii\ spectral  ranges closely follows previous 
work \citep{borosongreen92,marzianietal03a} and has been discussed in 
\citetalias{marzianietal13}. We only mention that emission blends near \mgii\ are 
mainly due to \feii\ which is modeled with the \feiiuv\ emission template provided by 
the last model in Table  1 in \citet{bruhweilerverner08}.  Main results presented in 
this paper were also tested with an additional empirical template produced by \citet{tsuzukietal06}. 
\fei\ emission was  detected in \citetalias{marzianietal13}, but  appears to be   fainter ({ along with any contribution due to Balmer continuum}) in B spectral types. 
Adopted ranges for  the multicomponent fitting analysis are 2600 \AA\ -- 3050 \AA\ and 2400--3100 \AA\ for Pop. A 
and B respectively. Known non-iron lines in this region include [{Ne}{\sc iv}] 2423.8,  [{O}{\sc ii}] 2471.0, 
semi-forbidden {Al}{\sc ii}] 2669.95, { \ion{O}{\c iii} 2672.04}, and most probably {He}{\sc i} 2945.11. 

\subsubsection{The ratio of the \mgii\ lines}
\label{ratio}

\begin{figure*}
\includegraphics[scale=0.6]{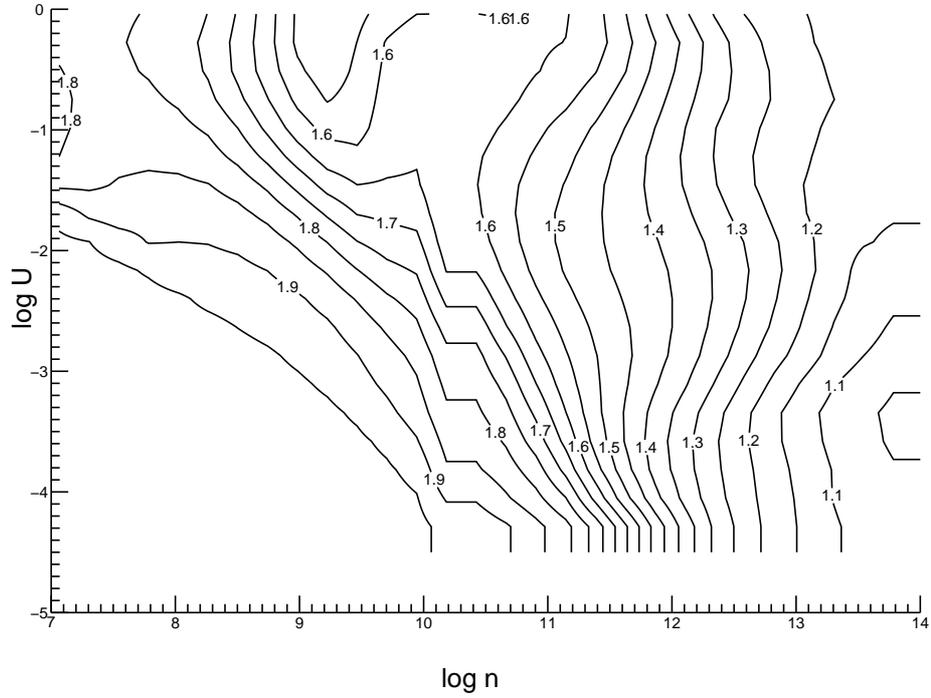}
\caption{Predicted intensity ratio of the doublet the Mg$^{+}$\ lines as a function of ionization parameter $U$ and hydrogen density \nh. Contour lines are drawn at steps of 0.1 from 1.0 to 2.0. \label{fig:mg}}
\end{figure*}

The intensity ratio of the \mgii\ doublet $\mathbb{R}$ = $I$(\ion{Mg}{\sc ii}$\lambda$2796.35) / $I$(\ion{Mg}{\sc ii}$\lambda$2803.53)\footnote{In this paper vacuum
wavelengths are used. Line identification includes the optical wavelength of traditional use.
For example, we refer to \oiii\ even if the reference wavelength is 5008.2 \AA.}
due to transitions
$^{2}P_{\frac{3}{2},\frac{1}{2}} \rightarrow ^{2}S_{\frac{1}{2}}$  should
be 2:1 in the pure optically thin case. In the case of infinite optical depth
the ratio should be 1:1 (i.e., if lines are fully thermalized). For intermediate conditions
the ratio depends on electron density, temperature, and column density.
Under standard assumptions the condition for thermalization { on the product ionization parameter times hydrogen density \nh}
is   $U$\nh$\gtsim$ 1.7 10$^{7}$ \cm3\ \citep{laoretal97a,hamannetal95}.  We therefore expect an $\mathbb{R}$\
value close to 1:1 for likely values of $U$\ and \nh\ in both the BLR and the  very broad line region (VBLR) associated with the VBC. An array of
{\sc cloudy} \citep{ferlandetal98} simulations predicts the intensity ratio
as a function of hydrogen density \nh\ and ionization parameter $U$\
(Fig. \ref{fig:mg}; a fixed hydrogen column density of $10^{24}$\cmq\ and
solar abundances are assumed). 

The individual lines are too broad and hence too blended to permit reliable
direct measurement. A-priori values must be assumed for BC and VBC (the
latter only for Pop. B sources) and the values could be different {\it in
lieu} of the different physical conditions expected for the BLR and the VBLR
\citep{marzianietal10}.   In the BLR,   $\log U = -2.75$, $\log $\nh = 12.50 yield  $\mathbb{R} \approx $ 1.3:1. The observed
ratio $\mathbb{R}$ in a typical Population A NLSy1 source (I Zw 1) is 1.2.
We assume a {\em standard} value of 1.25, noting that a small change of
$\pm 0.05$ in the doublet ratio does not affect line decomposition and
FWHM measurement. 

The intensity of \mgii\ is difficult  to compute by photoionization codes
because the line is primarily emitted in the partially ionized zone,
where a local approximation (i.e., via a local mean escape probability)
treatment of radiation transfer is probably inadequate. It is  possible that
the \mgii\ lines become fully thermalized at the extreme optical depth of the
LIL emitting part of the BLR, justifying the assumptions of a 1:1 ratio. In the following we have therefore
considered both cases. Mock \mgii\ profiles were computed for 1:25 and 1:1
cases assuming a wide range of line FWHM. The peak wavelengths  used as a
reference whenever the doublet is  treated as unresolved,  are 2800.1 \AA\
and 2799.4 \AA\ for the   1:1,1.25:1 cases respectively.   These values are found
to be almost independent of FWHM.  We assumed that the doublet wavelength cannot be shorter  than 2799.1 \AA, the value obtained for the case  $\mathbb{R} =$1.5:1.    Values of $\mathbb{R} \ga 1.5 $\ are  unlikely  for the physical conditions expected within the BLR (Fig. \ref{fig:mg}). 

In the VBLR,  $\mathbb{R} \approx$ 1.5  is   appropriate for
$-1 \la \log U \la -0.5 $ and for high density values (\nh$ \ga 10^{11}$\cm3), and only values  $\mathbb{R} \ga 1.7 $\ are  unlikely. Considering however that the doublet component  separation is $\Delta\lambda \approx 8$ \AA\ $\ll$ FWHM(VBC), the VBC can be treated as a single line. Under this assumption  the doublet ratio is indeterminate.

\section{Results}
\label{results}

\subsection{Broad Line Profile Shapes }

{ Composite spectra for Pop. B sources are shown in Fig. \ref{fig:hbmgb}, and Fig. \ref{fig:hbmgb2}, after continuum subtraction (Pop. A median spectra were shown in \citetalias{marzianietal13}).}
Table \ref{tab:prof} reports FWHM, asymmetry index A.I., kurtosis $\kappa$ and centroid at
fractional intensity measures
for the  \hb\ and \mgii\ broad profiles (i.e. after narrow components and \feii\
subtraction but before broad/very broad component   decomposition) of  both Pop. A and B spectral types. A.I.,    $\kappa$ and centroids  are  defined by Eqs. 1 -- 3 in
\citet{zamfiretal10}. Note that A.I. is
defined at $\frac{1}{4}$\ maximum intensity  and with respect to line peak
while centroids are relative to source
rest frame. These parameter values provide an empirical description of the
profile without any model assumption.
Reported uncertainties are at 2$\sigma$\ confidence level and have been
computed measuring the effect on
the profile parameters of changes in continuum placement (always less than
2\%, due to the extremely high s/n).
Note that full profile measures for \mgii\ doublet assumed an unresolved
single line. Width measures for \mgii\
can be converted to single component width by subtracting $300 $ \kms\
\citep{trakhtenbrotnetzer12}.
Zero point wavelengths set by \oii\ and \hb\ narrow component agree within an rms of 10
\kms. Uncertainties were estimated by propagating the zero point rms, the scatter associated with SDSS wavelength calibration and the
uncertainty of continuum placement. 

{ The most important results from Table
\ref{tab:prof} are the difference in line profile properties for Pop. A and B. FWHM differences  among Pop. B spectral types arise from the definition of 4DE1 and how it is binned.  Pop. A sources show broader, more symmetric and less shifted
\hb\ profiles than \mgii\ with the exception of bins A3 and A4 where a blue asymmetry
is observed. \mgii\ and \hb\  both show a symmetric
unshifted profile in bins A1 and A2.  The broad \mgii\ profile appears symmetric but  blueshifted by $\approx$--300 \kms\ in bins A3 and A4 and these are the only sources  where FWHM \mgii\ exceeds FWHM \hb.  This has been seen before and interpreted as  evidence for outflows in the highest \lledd\ A3 and A4 sources \citepalias{marzianietal13}, possibly because of a distinct LIL component  connected to the blue shift observed in HILs like \civ.  }


{ Pop. B sources are largely represented by bins B1 and B1$^+$.   Figure \ref{fig:hbmgb} confirms previous claims \citep{wangetal09} that FWHM \hb\ is systematically broader than  FWHM \mgii\ by approximately 20\%. This appears to be true for all B bins as well as bins A1 and A2. Both \hb\ and \mgii\ Pop. B profiles show median red asymmetries; \mgii\ show perhaps slighly weaker asymmetries. Centroid shifts are redward and of larger amplitude in \hb\ than \mgii. The red asymmetry in the lower half of the  profile was previously interpreted, in the case of \hb, as a signature of the VBC \citep{sulenticetal00c,sulenticetal02}.  Apparently \mgii\  shows a
(weaker) VBC component. Much of the excess \hb\ profile width may be due
to a stronger VBC component. In other words, the measures on line profiles of Pop. B reported in Tab. \ref{tab:prof}    suggest that  \hb\ is more strongly affected than \mgii\ by non-virial motions especially toward the line base. However, the VBC cannot be a full explanation for the difference in line width between \mgii\ and \hb\ because  A1 and A2 (where no VBC is present) show the same effect. }

Somewhat higher kurtosis values are found for broader Pop. B sources since Pop. A profiles are more sharply peaked \citep[c.f.][see \S \ref{origin} for interpretation]{kollatschnyzetzl13}. The kurtosis index is consistent with a single Gaussian profile ($\kappa \approx 0.44$) only for spectral type B1$^{++}$\ for both \hb\ and \mgii.  B1$^{++}$ profiles are rare, very broad and complex sometimes showing double peaked lines although not in our sample.

\subsection{Multicomponent Analysis}

\begin{figure*}
\includegraphics[scale=0.35]{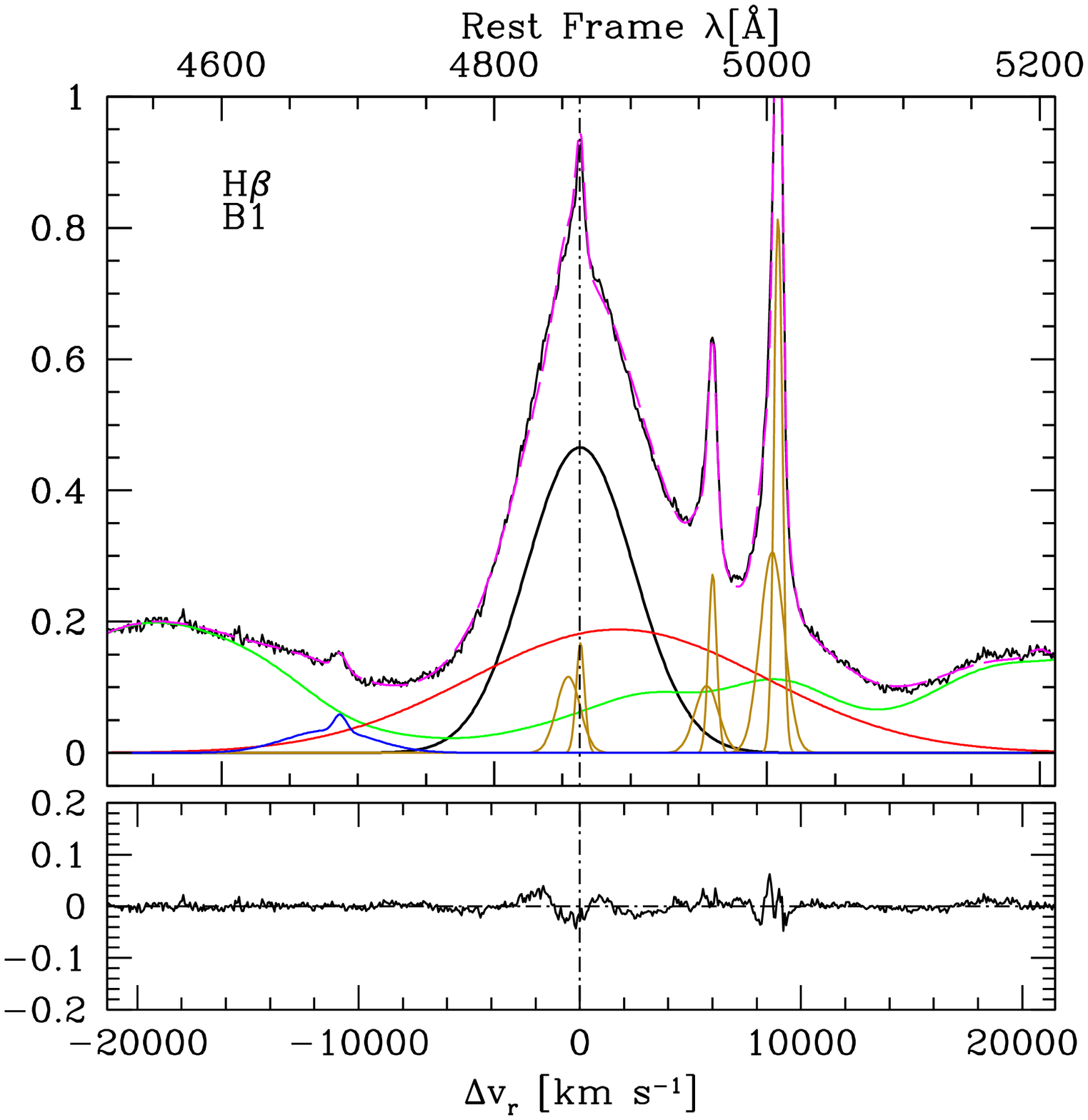}
\includegraphics[scale=0.35]{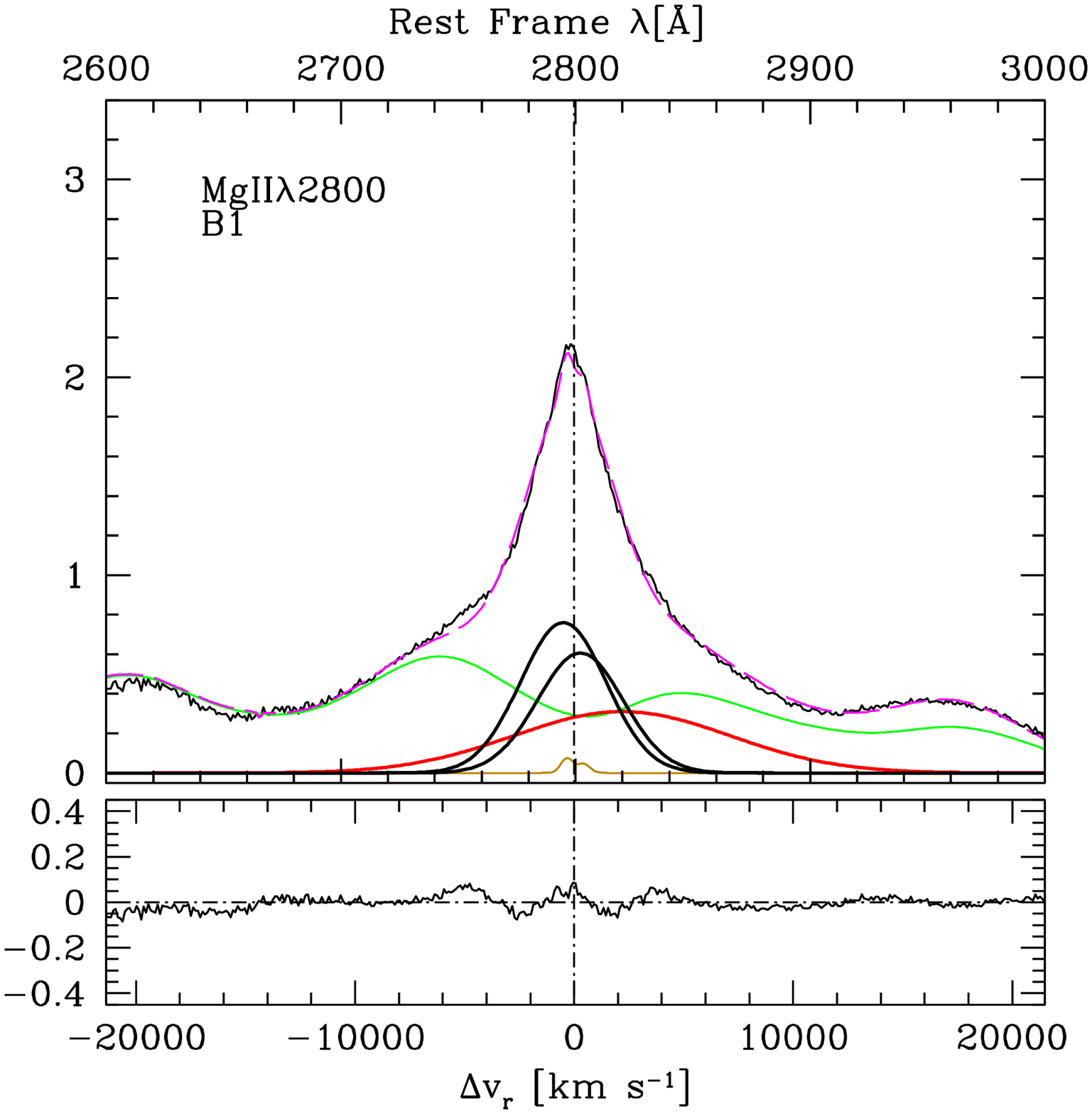}\\
\includegraphics[scale=0.35]{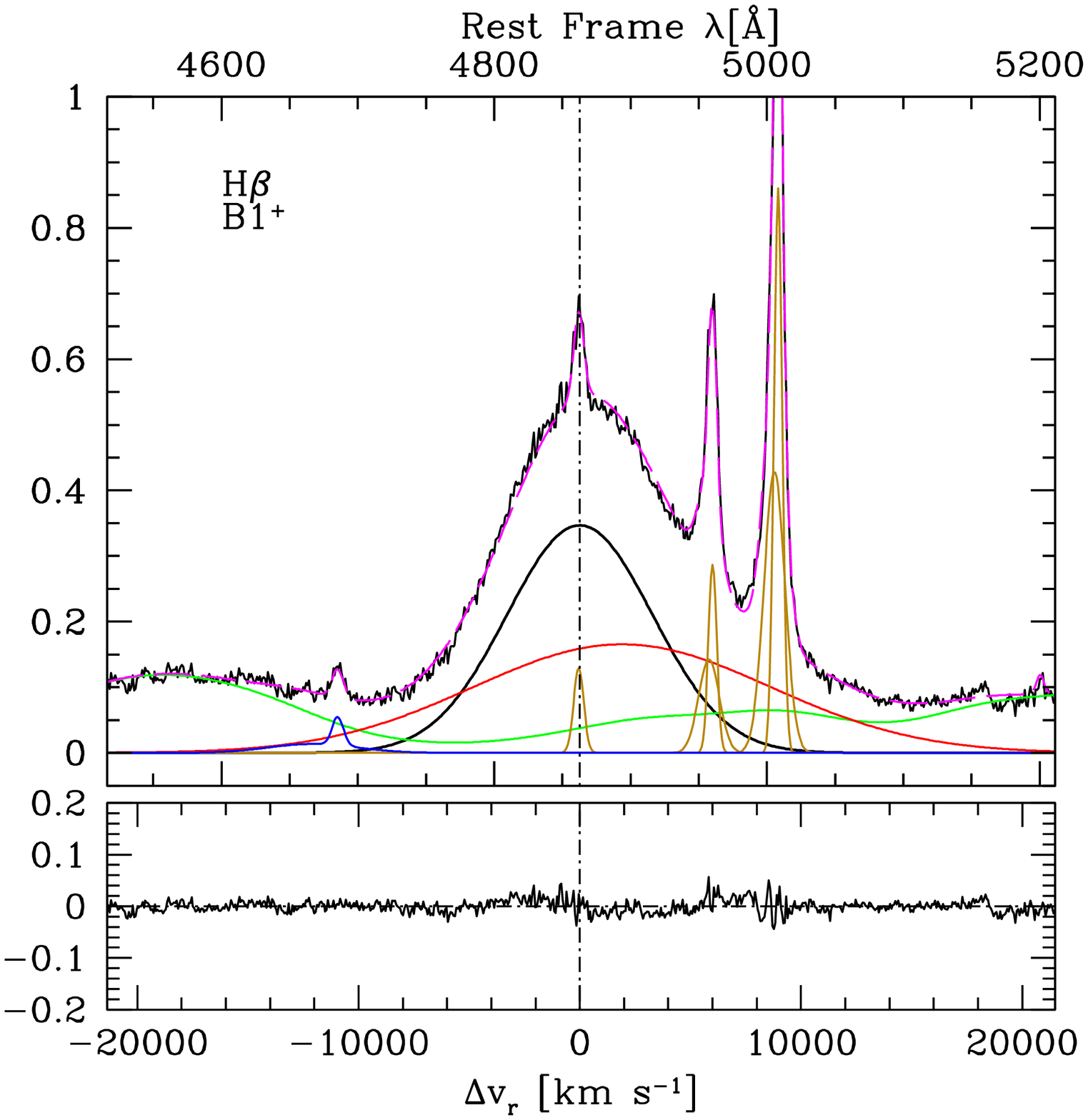}
\includegraphics[scale=0.35]{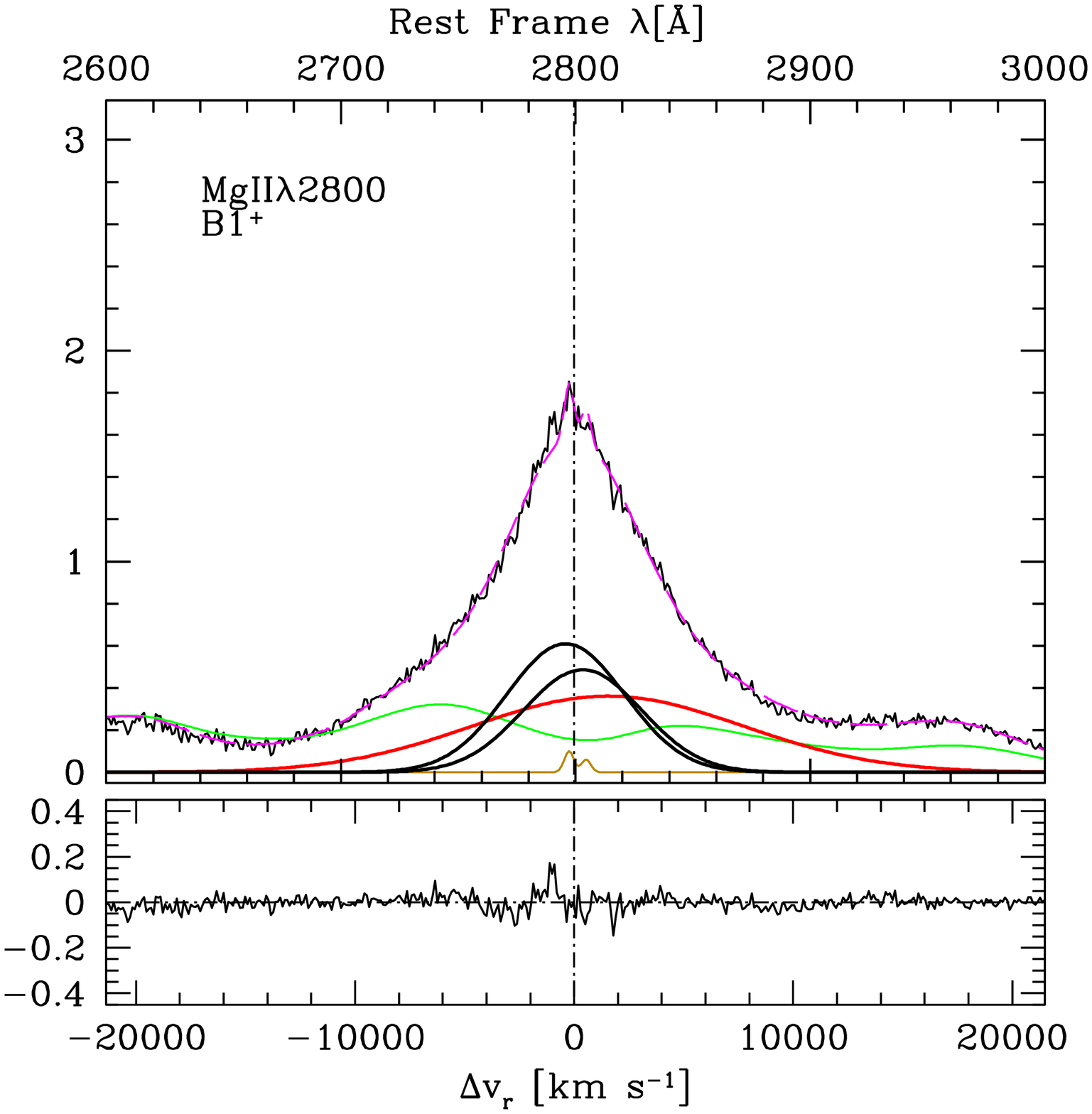}\\
\includegraphics[scale=0.35]{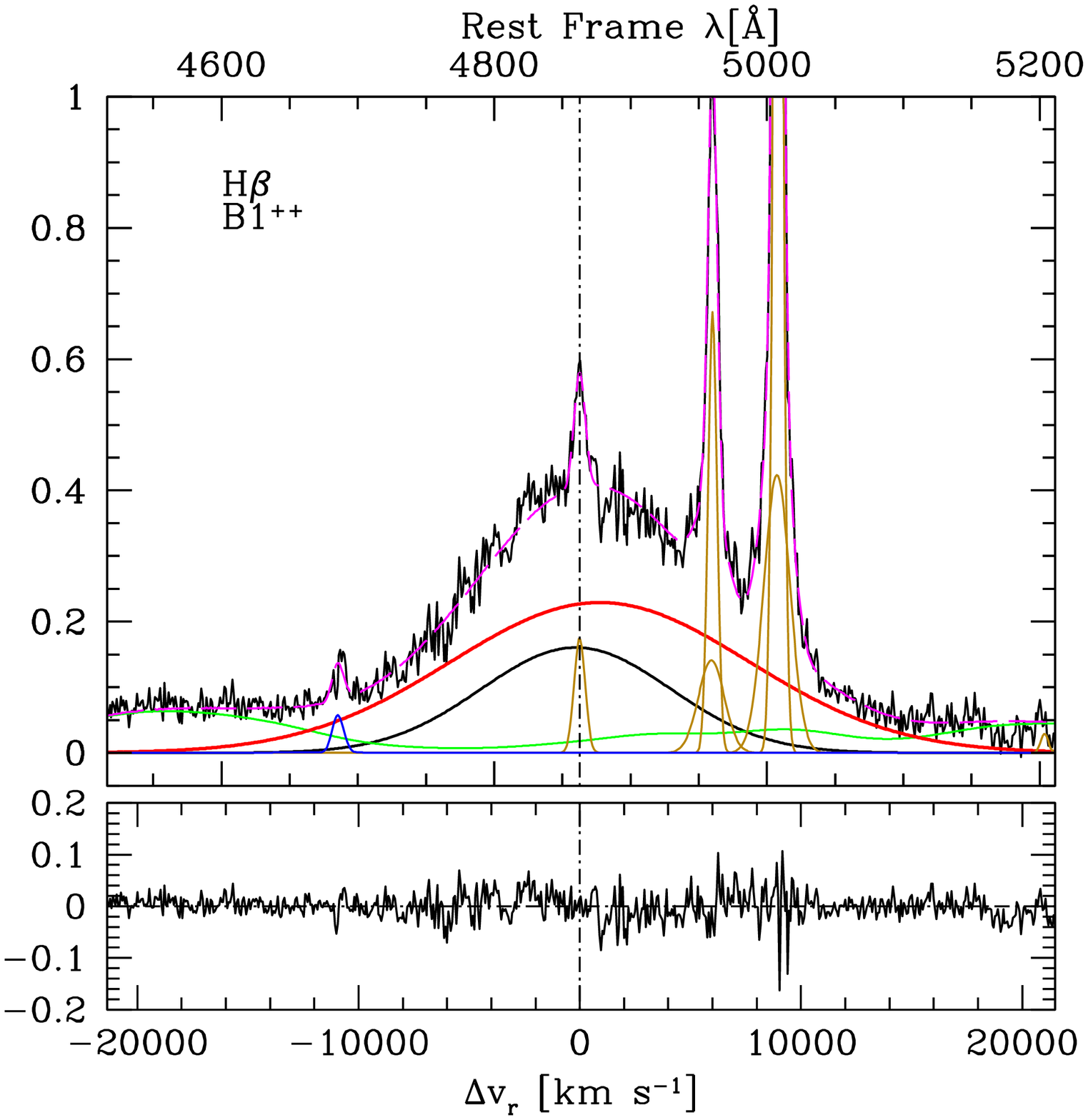}
\includegraphics[scale=0.35]{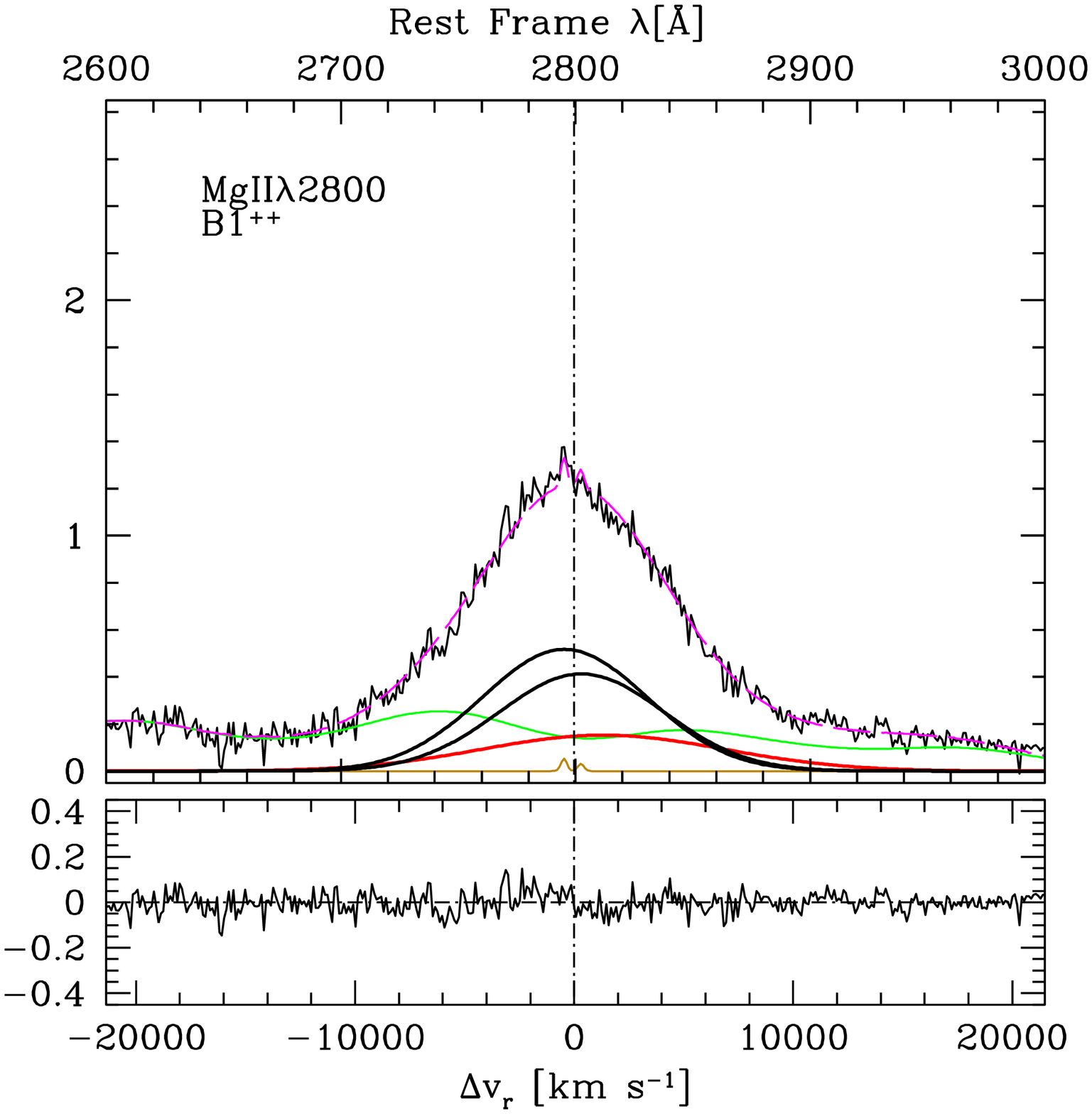}
\caption{Spectra of \hb\ (left panels) and of \mgii\ (right panels) for spectral types B1 to B1$^{++}$\ (from top to bottom). The horizontal scales are in rest frame wavelength [\AA] or radial velocity, with the origin set at the rest frame wavelength. The black lines shows the original, continuum-subtracted spectrum, while the dashed magenta line the model with all emission line components.  The thick black line is the BC; the thick red line shows the VBC. The green lines trace the \feiiopt\ and \feiiuv\ contribution, and the gold lines various contributions associated to the NLR (\hbnc, \oiiiopt, and \mgii\ narrow component when appropriate). {\bf Excess emission at $\lambda \gtsim 2900$\AA\ visible in this and in the next two Figures is  likely due to a combination of \fei\ and Balmer continuum emission.} \label{fig:hbmgb}}
\end{figure*}

\begin{figure*}
\includegraphics[scale=0.35]{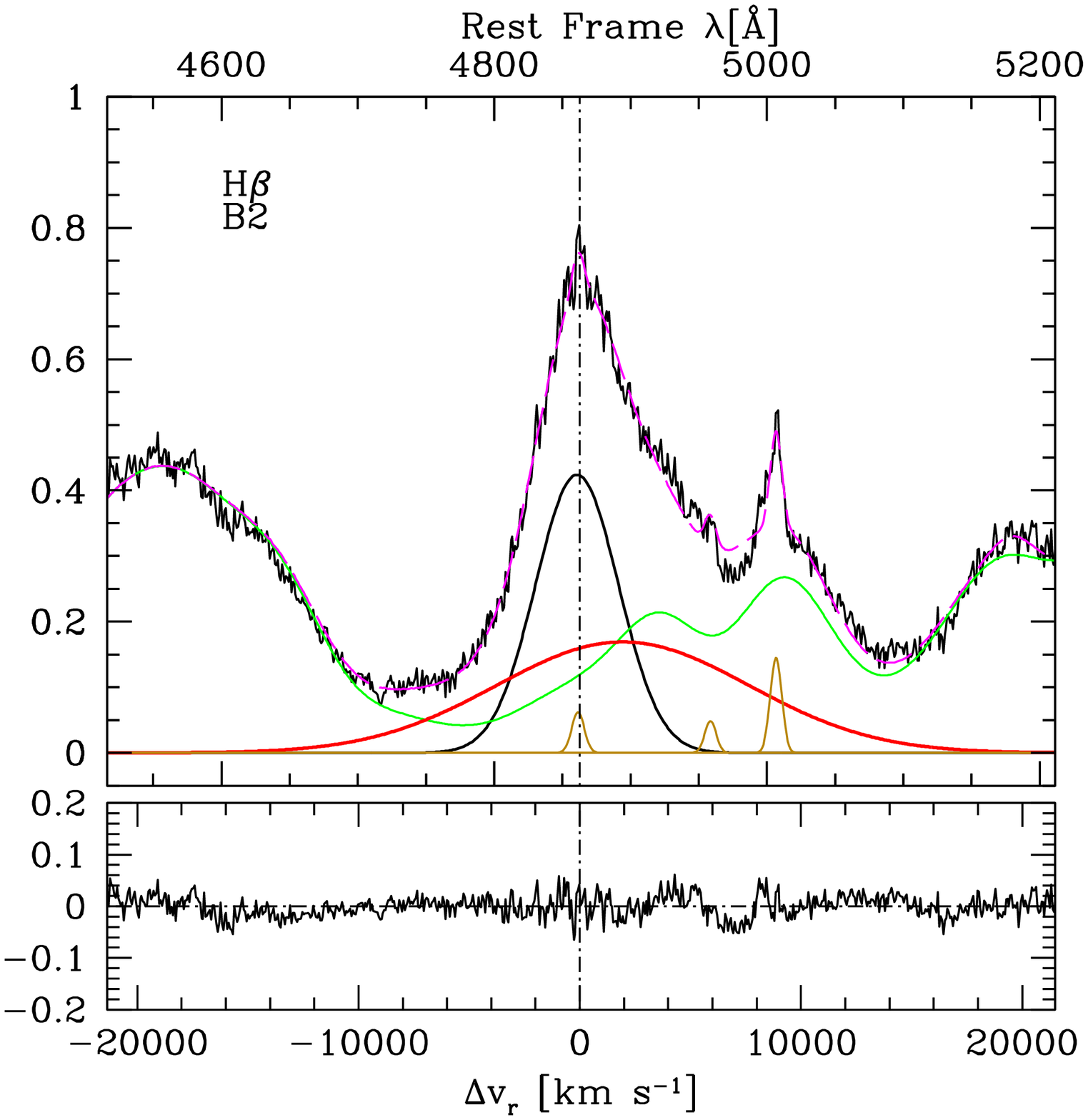}
\includegraphics[scale=0.35]{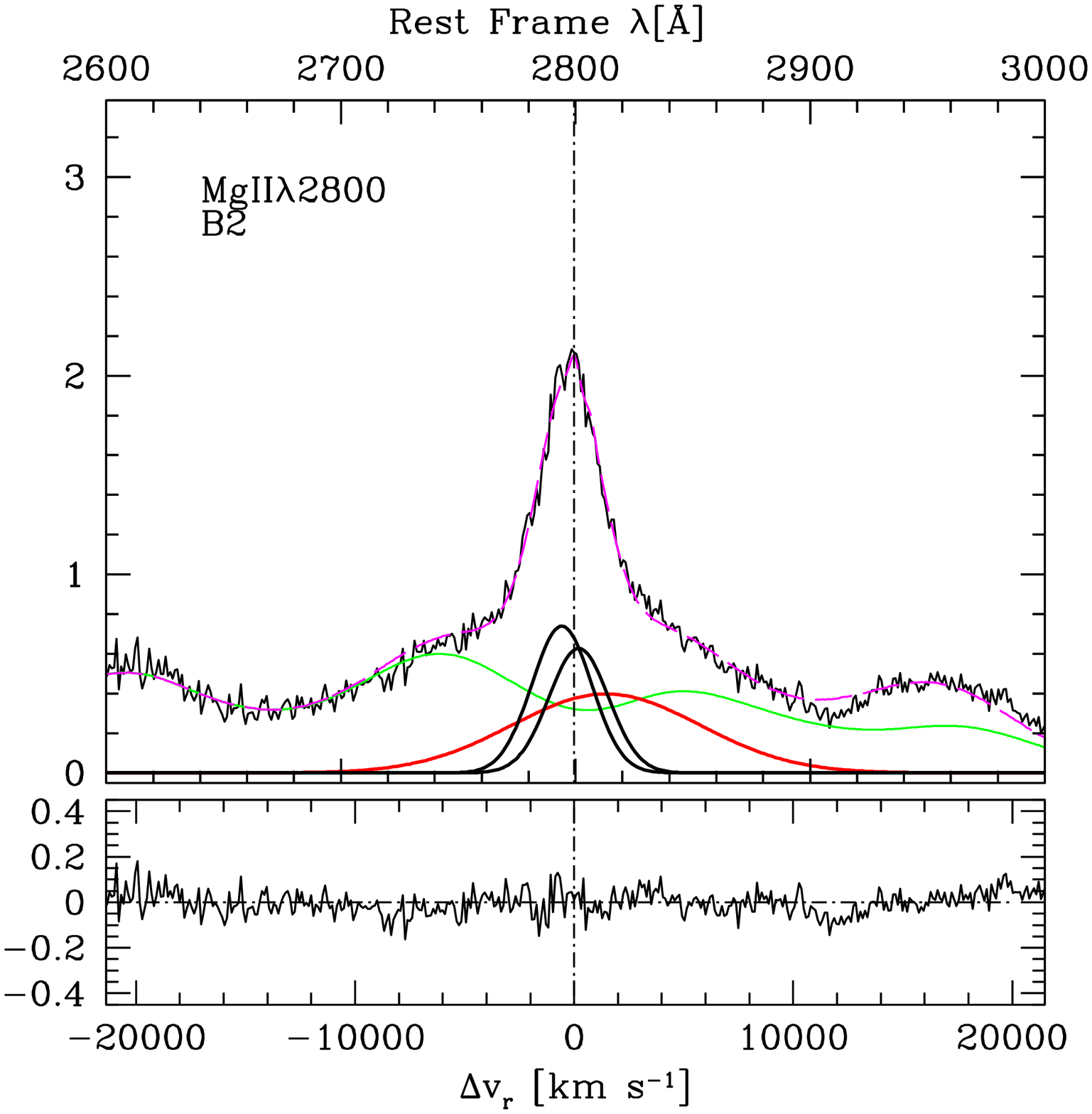}
\caption{Spectra of \hb\ (left) and of \mgii\ (right) for spectral type B2. Meaning of   symbols/line colors is the same of the previous Figure.\label{fig:hbmgb2}}
\end{figure*}

{\sc iraf specfit}  \citep{kriss94} was used to model and decompose line
blends and components
in Pop. B composites as previously reported  for Pop. A composites
\citepalias{marzianietal13}.
We assumed a weak \mgii\ narrow component in Pop. B composites (see \citealt{wangetal09}
and references therein)
constrained to not exceed the ratio \mgii/\hbnc $\approx$ 2 (appropriate
for $\log U \sim -2 $ and
$\log $\nh$\approx 4 - 5$). A double Gaussian model involving an unshifted
BC and redshifted VBC
was used for fits to both lines following previous work on \hb\
(\citealt{marzianietal09};  Fig. \ref{fig:hbmgb}, and Fig. \ref{fig:hbmgb2}).
In contrast to Pop. B, the
majority of Pop. A profiles
were well fit by an (almost) unshifted symmetric Lorentz function. Table
\ref{tab:specfit} presents
BC and VBC measures (intensity, peak shift and FWHM) for \hb\ and \mgii\
for B spectral type and
for the source \object{PG 1201+436}.  Uncertainties are $2 \sigma$\
confidence level. Uncertainties of peak shifts were computed by quadratically propagating
errors on zero point,
wavelength calibration rms and  uncertainty provided by the fitting
routine \citepalias{marzianietal13}.
Uncertainties reported for the VBC should be considered as formal errors
and are computed following
the assumption that that the best fit is the correct one and that the
functional form describing the
line component is also correct. The VBC is a broad component with poorly
constrained centroid
and width. More realistic uncertainties are estimated around 20\%\ for
both shift and width measures.

Our operating assumption is that unobscured gas in virial motion should
give rise to a reasonably
symmetric/unshifted line profile. The BC component is therefore the virial
estimator that we are trying
to isolate. The use of a double Gaussian model in {\sc specfit} was
motivated by the inflected
appearance of the \hb\ profile (this is evident in  the B1 panel of Fig.
\ref{fig:hbmgb}).
Decomposition of the BC/VBC blend was the most uncertain part of the
modeling. We derive the
ratio of BC / VBC intensity for the two lines from the fit in the B1, B2 and B1$^+$ cases. 
However, we do not know
the shapes of the individual
components. A highly redshifted VBC is unlikely to show a symmetric
Gaussian profile. Past work has
focussed on \hb\ and has used very high s/n composites like the ones
presented here. Past work involving
BC+VBC decomposition \citep{sulenticetal02,marzianietal09} is limited and
shows FWHM \hb\ BC  and shift
values in the range 4000-4800 \kms\ and zero \kms, respectively, with the
relative strength of the VBC possibly increasing with source luminosity.

Modelling is clearly more ambiguous in sources with
broader multicomponent profiles. The \mgii\ profile in Pop. B sources also
deviates from a single Gaussian showing a narrower core and more prominent wings. In other words it is  more similar to a Lorentz/Voigt profile that can be approximated by the sum of the Gaussians. An attempt to fit \mgii\ in the B1 composite
with a single Lorentzian will reproduce the observed red wing but model
the blue one with  a significant excess. The sum of two Gaussians, with the broadest one slightly redshifted provides a satisfactory fit (Fig. \ref{fig:hbmgb}).
The introduction of a small shift in the broader Gaussian accounts for the
excess flux in the red wing, prominent in \hb\ but also detected in \mgii. Nonetheless the assumption that the VBC is a shifted symmetric Gaussian is unlikely to be correct for such a redshifted component: non-virial motions are likely to strongly
affect the VBC.  We currently have no empirical basis for fitting the VBC
with more complex models. Our best current fits suggest that the \mgii/\hb\ intensity ratio differs for the VBC and BC: with \mgii/\hb $\gtsim$ 1.5
for the BC and \mgii/\hb $\ltsim$ 1 for the VBC. This result is consistent
with the idea that the VBC arises in a higher-ionization VBLR near the inner edge of the BLR \citep{koristagoad04}.

\begin{figure*}
\includegraphics[scale=0.35]{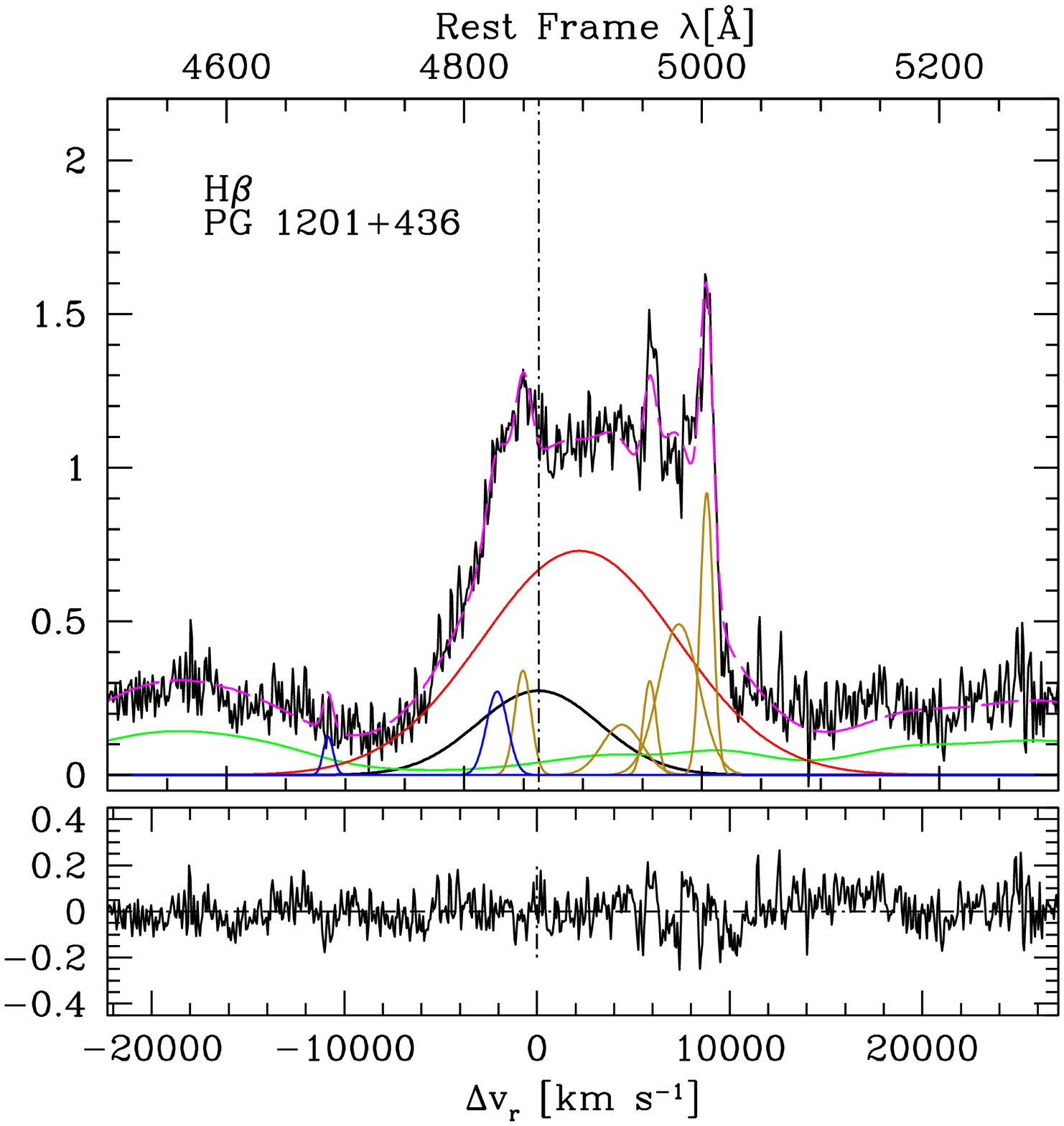}
\includegraphics[scale=0.35]{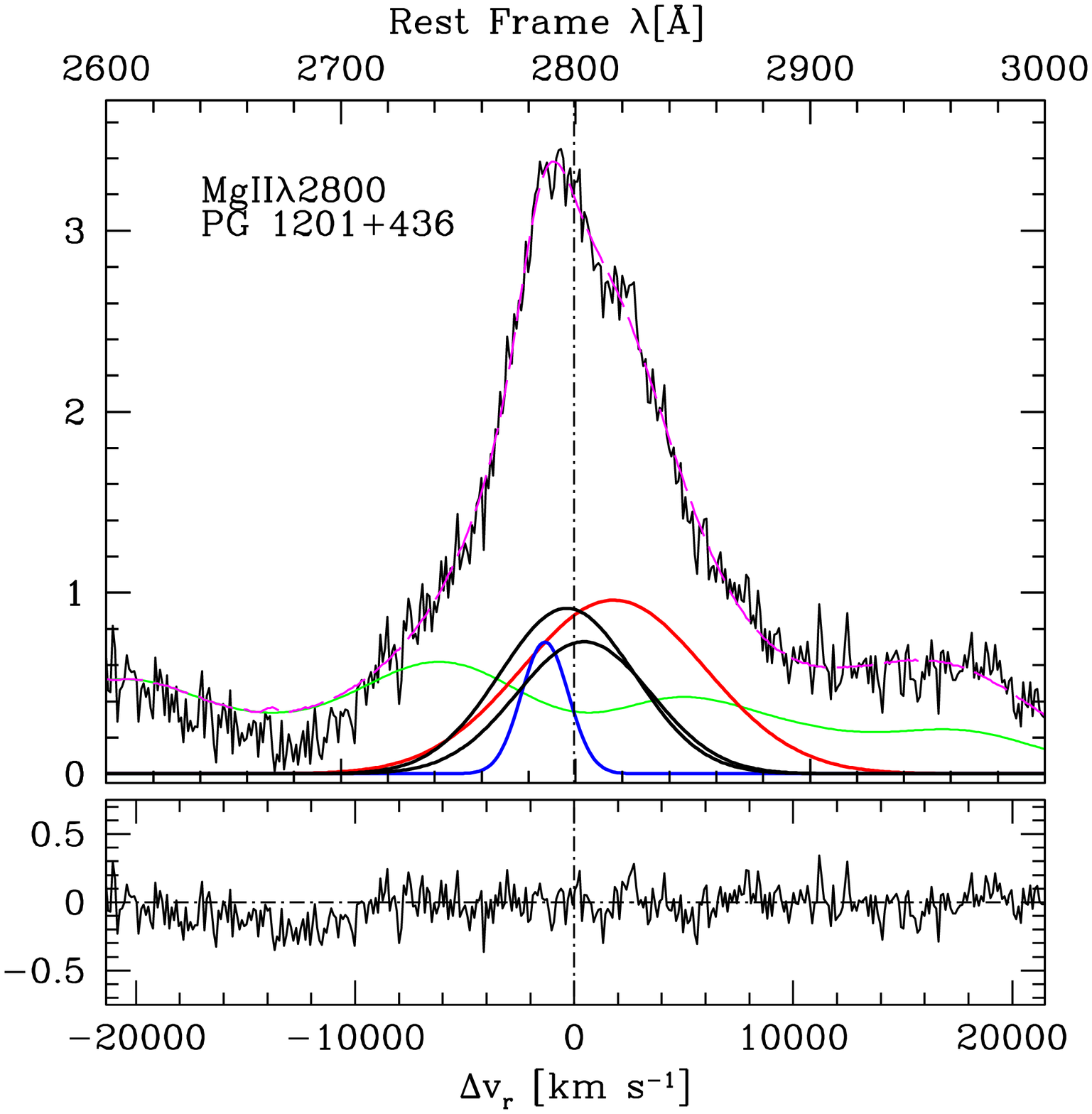}
\caption{Spectrum of \hb\ (left panels) and of \mgii\ (right panels) for the  B1$^{++}$\ source PG 1201+436. Color coding is the same as for Fig. \ref{fig:hbmgb} save for the blue line that traces a blueshifted peak visible in both \hb\ and \mgii\ (\S \ref{results} and \ref{origin}).  Horizontal scale is rest frame wavelength or radial velocity shift from rest frame.  Vertical scale is specific flux in units of 10$^{-15}$ erg s$^{-1}$ cm$^{-2}$ \AA$^{-1}$. \label{fig:pg1201}}
\end{figure*}

Modeling the VBC as a distinct component is further justified by the
discovery of sources where the \hb\ profile is dominated by the VBC with little or no detected BC
(e.g. \object{PG 1416--129} in \citet{sulenticetal00c} and 3C110 in
\citet{marzianietal10}).  In the course of assembling the sample under
consideration in this study we found a
few more VBC dominated sources. The best example involves \object{PG
1201+436} whose \hb\ and \mgii\ profiles are shown
in Fig. \ref{fig:pg1201}. Most of the broad redshifted \hb\ profile is 
almost certainly VBC with little or no BC emission.
The most straightforward interpretation of the \mgii\ profile is,
conversely, BC dominated with much weaker VBC emission.
We can use these considerations to cautiously guide {\sc specfit} in
modelling of \mgii\ in  bin B sources with the broadest profiles i.e.,
 B1$^{++}$. B1$^{++}$\  sources however constitute less than 3\%\
of our sample. They are however relevant if for no other reason that bin
B1$^{++}$\  shows the largest RL fraction in the 4DE1 sequence
(\citealt{zamfiretal08} and Tab. \ref{tab:numbers}). This is also the
domain of  some of the best known sources that show double-peaked profiles
\citep{eracleoushalpern03} although most do not. A BC+VBC profile with
peaks near zero + redshift is not a
double peaked profile.The  fit for B1$^{++}$\ shown in Figure 3 assumes a
BC-dominated \mgii\ profile and attempts to maximize \mgii\ BC intensity
and width. It is interesting that the resulting \mgii\ profile: 1) 
remains less broad than \hbbc\
and 2) requires a VBC component to produce an acceptable fit.

Results  detailed in  Table \ref{tab:specfit}  and Figures \ref{fig:hbmgb}
+ \ref{fig:hbmgb2} indicate that in Pop. B sources
\mgii\ mimics \hb\ by showing a VBC component. They also show that the VBC
has a stronger effect on FWHM \hb\ when measuring the full profile. In
this sense FWHM \mgii\ BC appears as a safer virial estimator than FWHM
\hb. BC measures are preferred as virial estimators for both lines with a
consistency correction factor taking into account  the fact that FWHM
\mgii\  (BC) is 20\% narrower than FWHM H$\beta$ for the majority of
sources  (B bins and bins A1+A2, 90 \%\ of our sample).  Nonetheless, all
measures reported in  Tab. \ref{tab:prof} and Table \ref{tab:specfit} for
all spectral types excluding A3 and A4 show  that the FWHM of {\em any}
\mgii\ component  is  narrower than the equivalent \hb\ component.
This result is robust since it also holds for the full profile. 

Tab. 
\ref{tab:specfit} indicates that B1 sources may show a  significant \mgii\
BC  blueshift,  with smaller amplitude in \kms\ than the one measured in
Pop. A. A small blueshift might also be present for B1$^{++}$. Part of the
BC blueshift might result  from modeling the VBC with a symmetric profile
and possibly over-subtracting flux from the red side of the BC.  It is
however unlikely that this accounts for the full effect since  a 
blueshift is also  measured for the full profile: a   blue-side excess is
visible in the median B1 \mgii\ profile shown in Fig. \ref{fig:hbmgb}.  
Tab. \ref{tab:prof} indicates a \mgii\ blueshift for spectral types B1 and
B2 with any blueshift in B1$^{++}$\  BC not detected by our
parameterization of the full profile. The relevance of the blue shifts in
Pop. B profiles will be further discussed in \S \ref{trends} and \S
\ref{discuss}.

\subsection{Radio Loud}

\begin{figure*}
\includegraphics[scale=0.225]{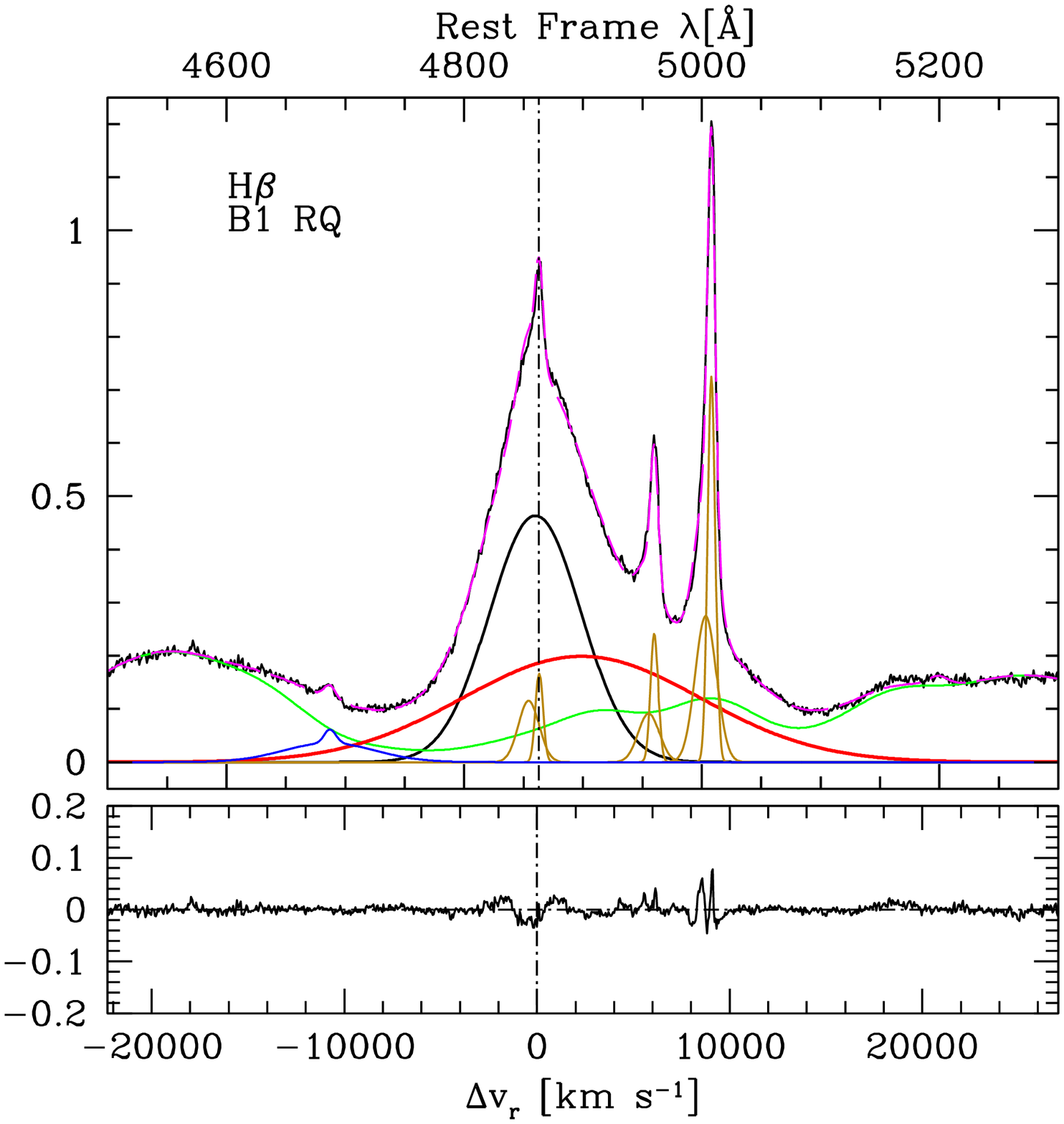}
\includegraphics[scale=0.225]{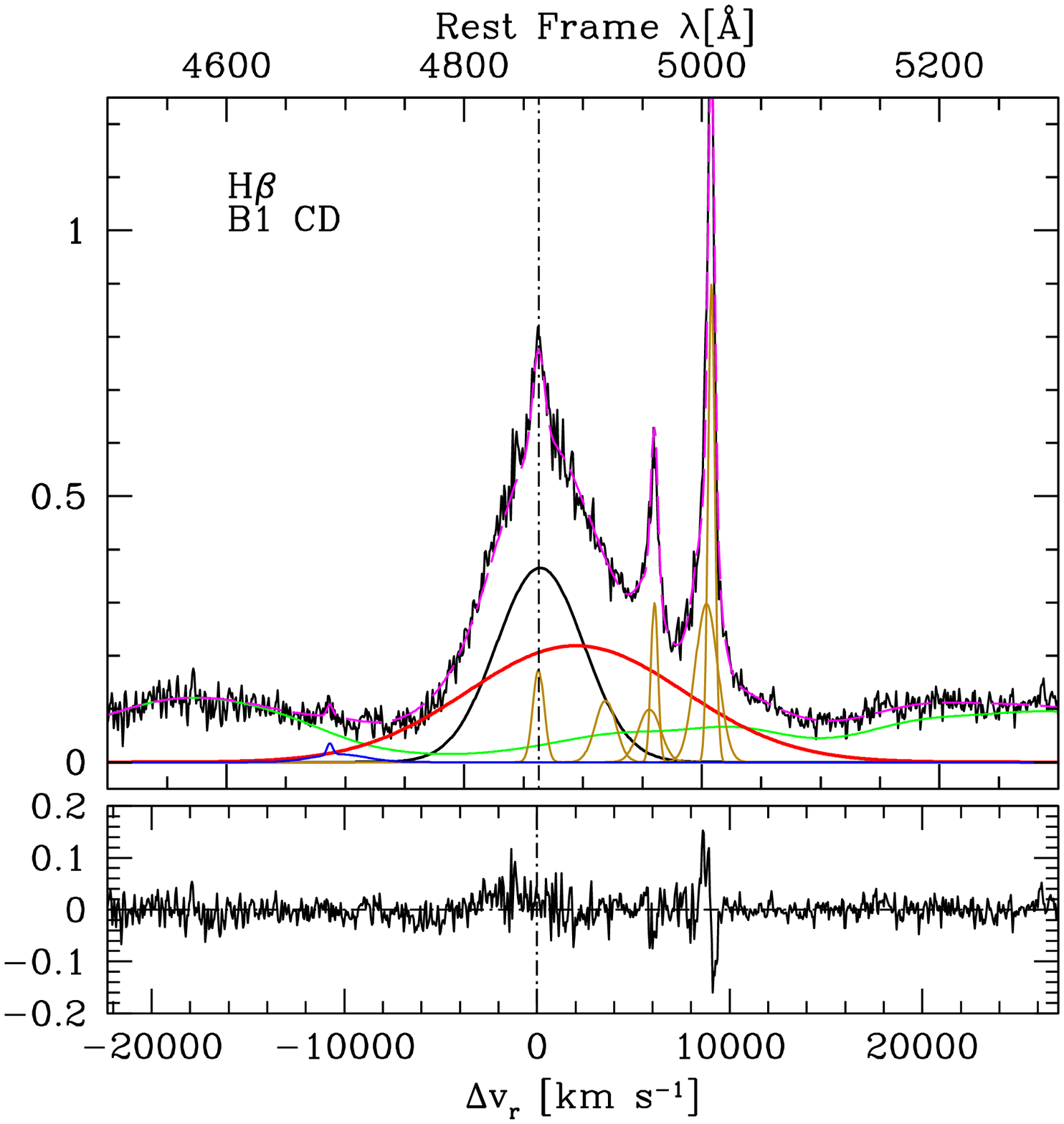}
\includegraphics[scale=0.225]{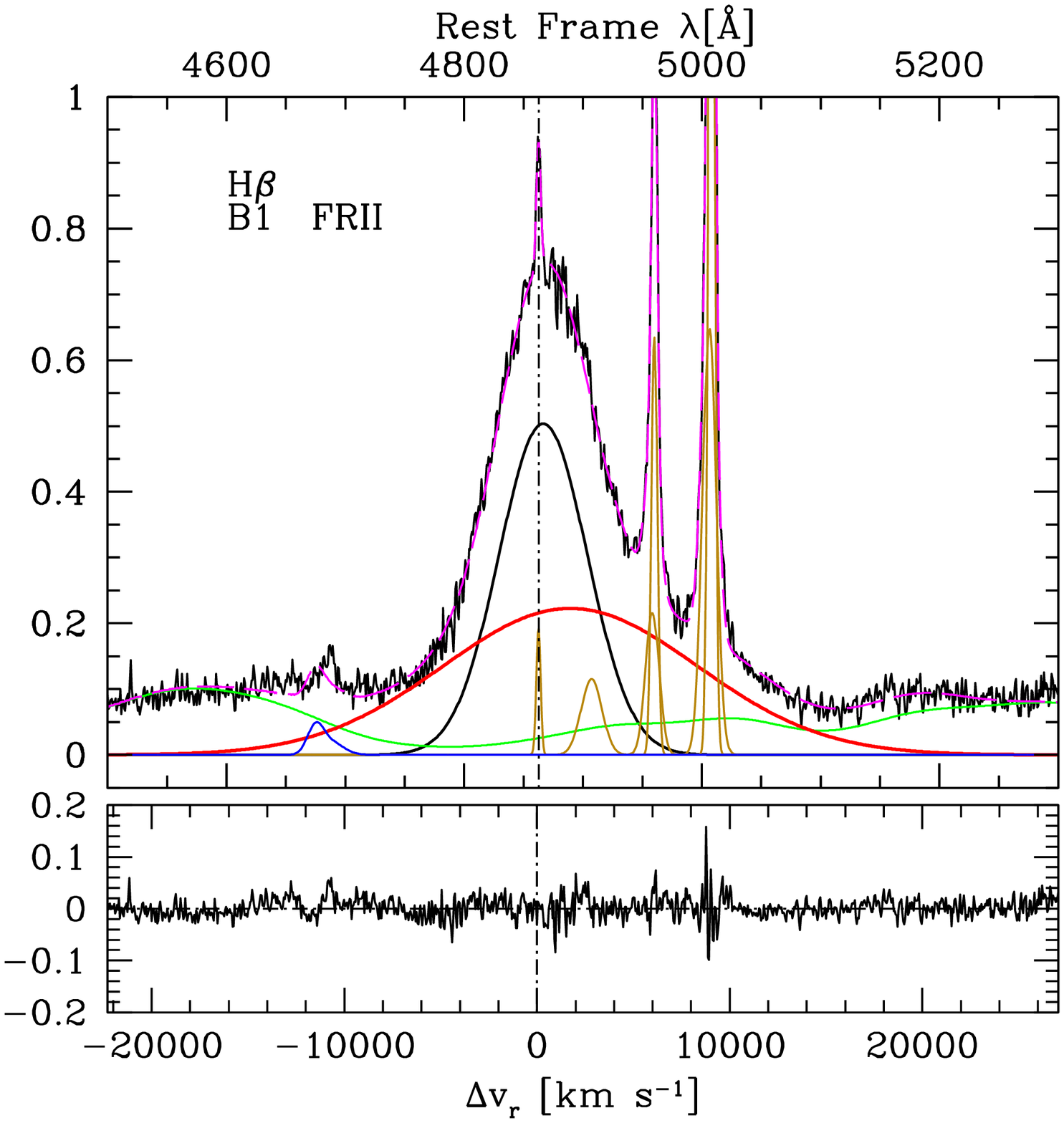}\\
\includegraphics[scale=0.225]{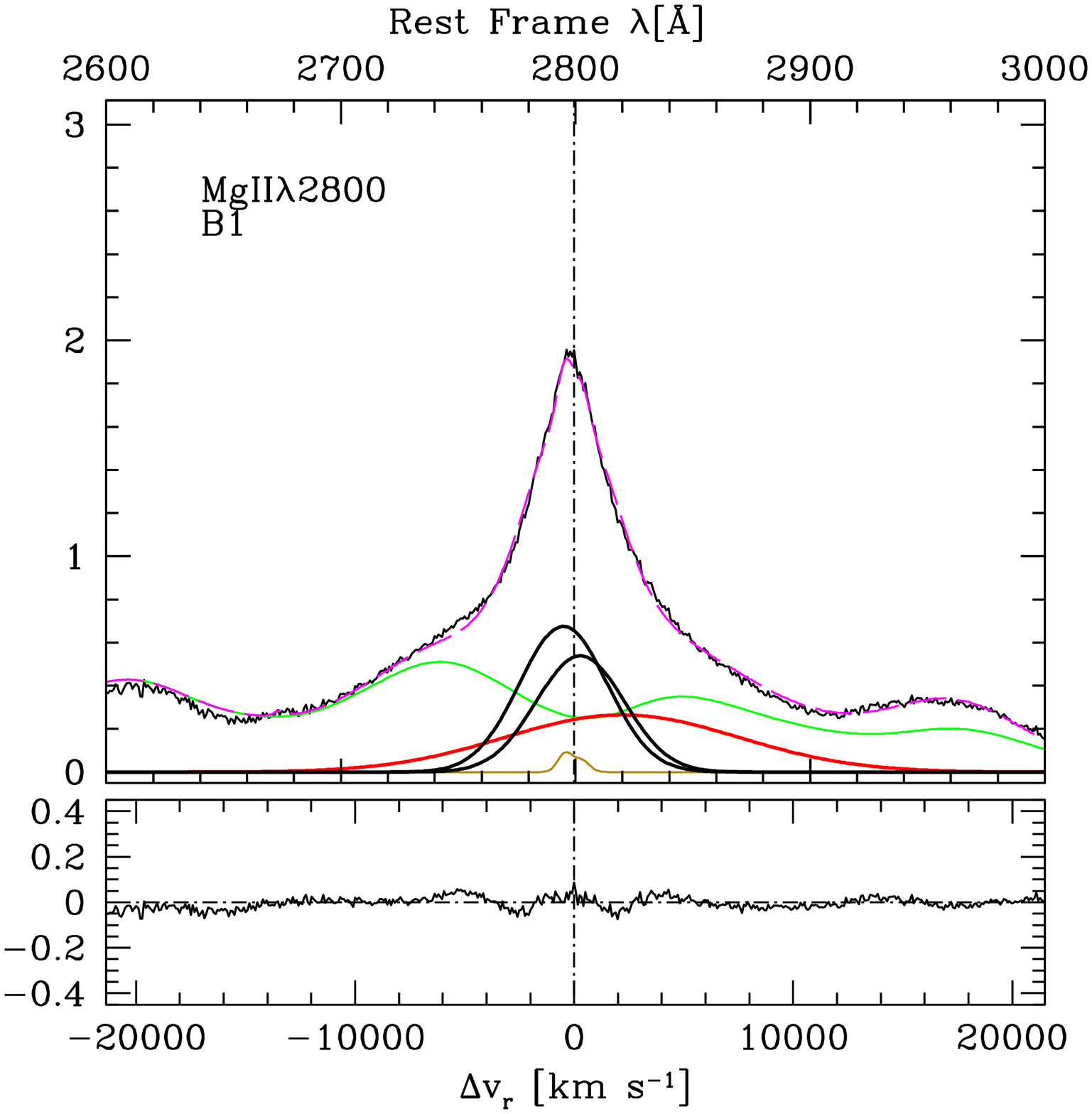}
\includegraphics[scale=0.225]{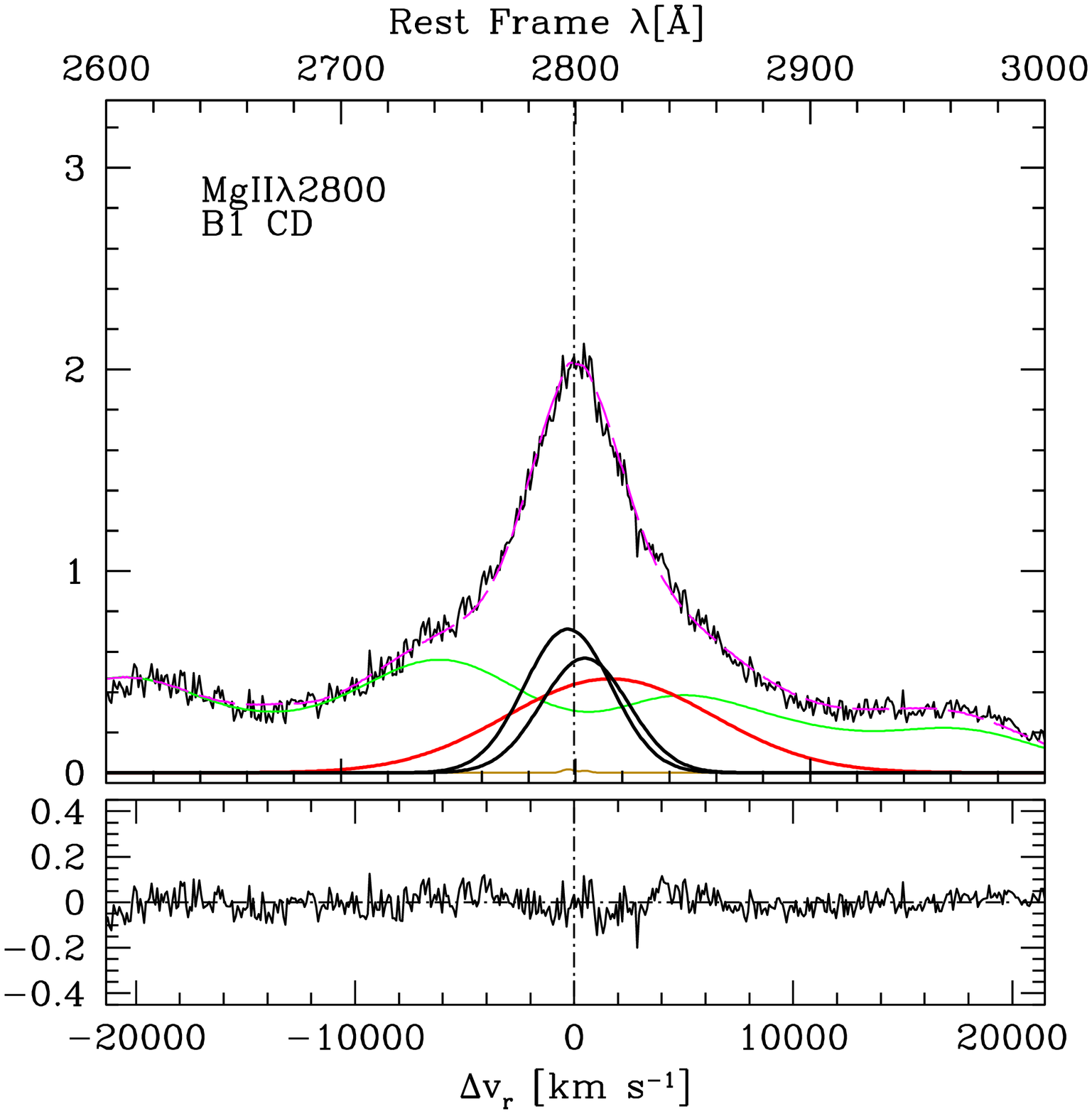}
\includegraphics[scale=0.2250]{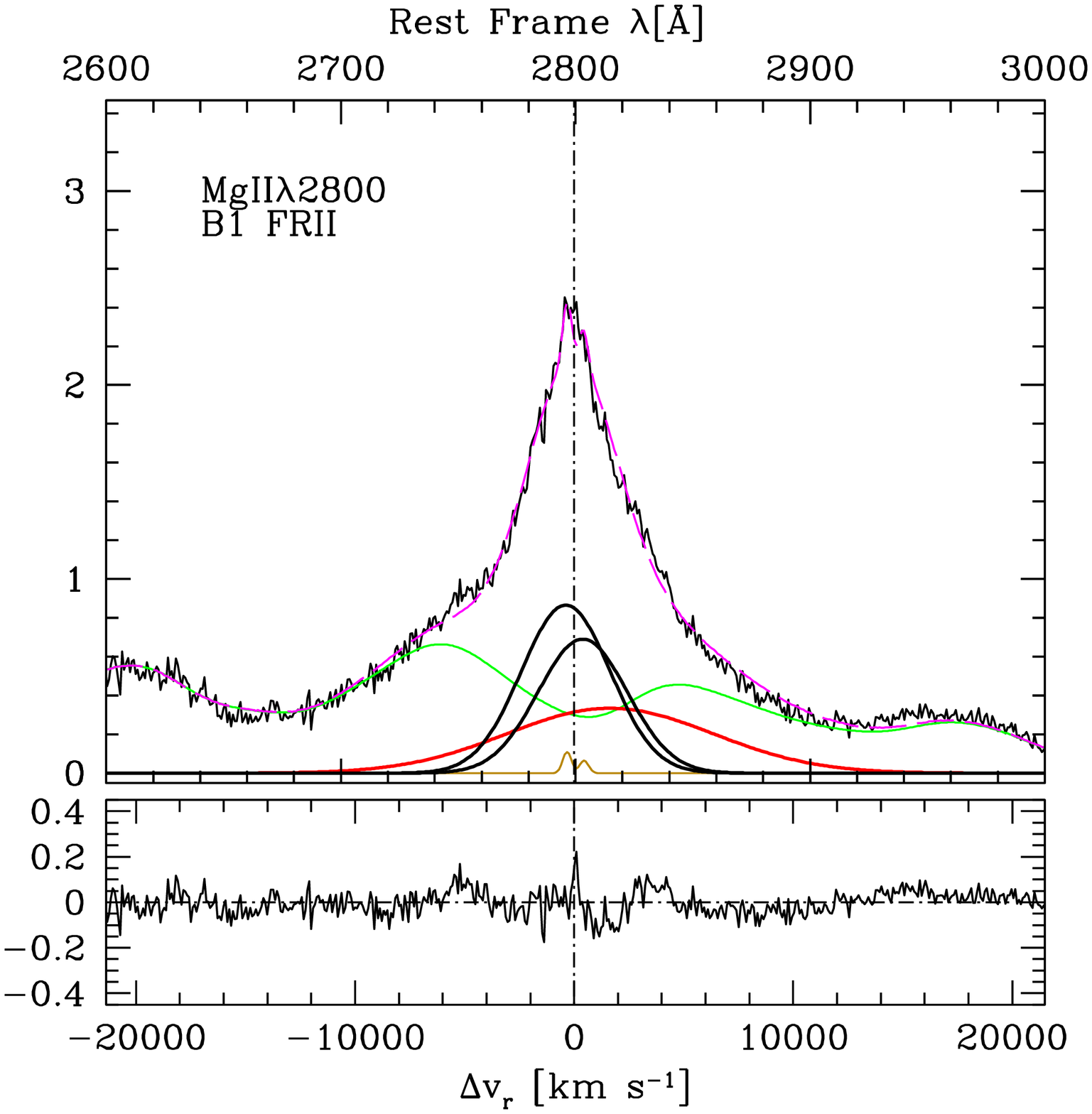}\\
\caption{{\bf Continuum subtracted spectral ranges of  \hb\ (top) and \mgii\ (bottom)  for RQ (left), CD (middle) and FRII (right)  median composites computed in the B1 bin. }Meaning of     symbols is the same of the previous Figures.\label{fig:hbmgrl}}
\end{figure*}

RL sources, both
CD and FRII (Fig. \ref{fig:hbmgrl}) B1 medians    show a   symmetric (FRII)
or redshifted (CD)  \mgii\ line core. The profiles resembles the typical B1 RQ and full sample profiles  but there is no hint of possible blue shifts.  A  shift to the red  is observed at 1/4 intensity level
but, again, the shift amplitude is less than for \hb. The redward
asymmetry is especially prominent in CD sources with a $\frac{1}{4}$\
centroid displacement reaching 600 \kms. CD and FRII sources do not show
any large difference in FWHM since they were both chosen to belong to
spectral type B1 by imposing  the same condition on  FWHM \hb. Their
interpretation will be discussed in \S \ref{rl}.

\subsection{Major trends}
\label{trends}

We focus on line shifts and widths because they are parameters relevant to
the credibility/definition of a virial broadening estimator. Comparison of
centroid shift values at different heights in a line give the best way to
evaluate asymmetry and line shift. 
The behavior of Pop. B is  different than Pop. A showing a redshift that is strongest in the line base
resulting in red asymmetric profiles. We assume that modeling and
subtraction of a VBC component will enable us to measure the width of a
symmetric and unshifted BC virial estimator. Pop A sources in bins A1 and
A2 yield a FWHM measure requiring no VBC correction. A correction is
needed only to make consistent virial estimate using FWHM \hb\ and FWHM
\mgii. In bins A3 and A4 the FWHM \hb\ measure is preferred after correction for a blue asymmetry.

Rather than the amplitude in \kms, a value normalized by line width 
provides a more  direct evaluation of the ``dynamical relevance'' of any
line shift. We can define the following  parameters

\begin{equation}
\label{ }
\delta(\frac{i}{4}) =
\frac{c(\frac{i}{4})}{\mathrm{FW}\frac{i}{4}\mathrm{M}},
i=1,2,3,
\end{equation}

\begin{figure*}
\includegraphics[scale=0.75]{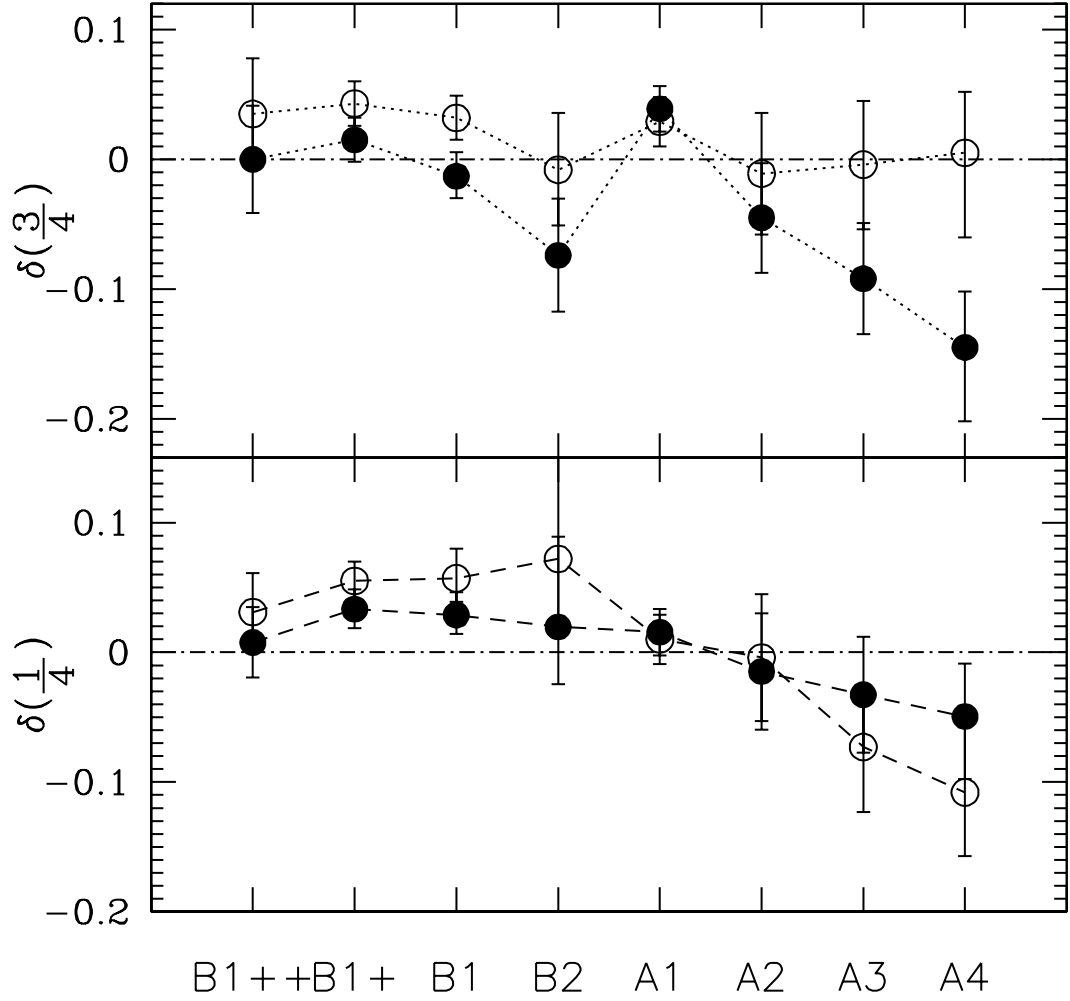}
\caption{Trends in Pop. A and B as a function of spectral type. The upper panel shows the normalized shift $\delta(\frac{3}{4})$, the lower  $\delta(\frac{1}{4})$. Filled symbols: \mgii; open symbols: \hb. \label{fig:trendsp} }
\end{figure*}

\begin{figure*}
\includegraphics[scale=0.75]{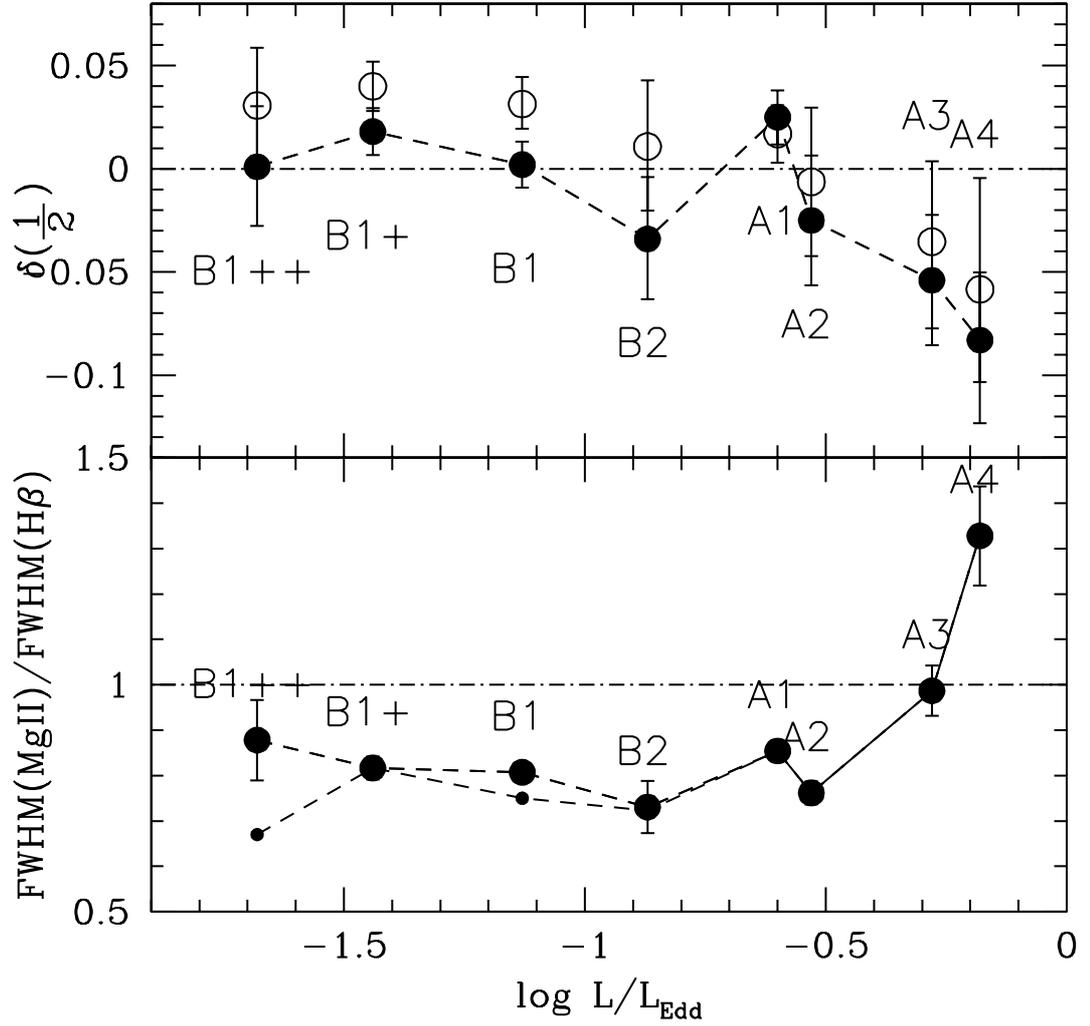}
\caption{Trends  as a function of Eddington ratio \lledd. Upper panel: ``dynamical relevance'' indicator $\delta(\frac{1}{2}) = c(\frac{1}{2})$/FWHM for the \mgii doublet broad component. Bottom panel: ratio FWHM(\mgii)/FWHM(\hb) for the broad components.  The smaller circles refer to measurements made including the VBC contribution of both lines. 
\label{fig:trendall}}
\end{figure*}

{ where the centroid $c(\frac{i}{4})$\ at fractional peak intensity
$\frac{i}{4}$\ has been normalized by the full width at the same
fractional peak intensity FW$\frac{i}{4}$M.} Fig. \ref{fig:trendsp} shows
the behavior of $\delta(\frac{3}{4})$ and of $\delta(\frac{1}{4})$\ as a
function of spectral type (error bars are obtained propagating
quadratically the uncertainties in Table. \ref{tab:prof}). A radial
velocity  \vr\ shift close to a line peak may represent a systematic
effect involving the entire emitting region   only if it is also seen at
$\frac{1}{2}$\ and $\frac{1}{4}$\ fractional intensities. Fig.
\ref{fig:trendsp} confirms that the B1 \mgii\ blueshift  is much less
relevant than shifts found in bins A3 and A4. We also note that in bin A3
and A4 the  shift measured for  the full \mgii\ profile is probably a
lower limit, since the profile can be interpreted as the sum of an
unshifted line + a blueshifted component (\S 3.3 of
\citetalias{marzianietal13}).  The line base might be affected by the
presence of additional kinematic components that, in our interpretation of
the line profiles, are the blueshifted component ({ blue} in
\citealt{marzianietal10}) and the VBC for Pop. A and B respectively.
Considering the $\frac{1}{4}$\ centroids as a function of spectral type,
only bins A3 and A4 show a relatively large \hb\  blueshift (the peaks are
almost unshifted). Bin A1 and A2 show symmetric profiles in both \hb\ and
\mgii.  The situation  changes in Pop. B where the B1 and B1$^{+}$\ \hb\
profiles are strongly red asymmetric with the B1$^{+}$\ centroid shift
reaching $\approx$ 800 \kms. Conversely the \mgii\ profile shows a more
symmetric shape. There is a significant shift at $\frac{1}{4}$\ intensity
level but its amplitude is $\sim$0.25 that of \hb. Given the large
FW$\frac{1}{4}$M for \mgii,  the $\delta$\ values are lower than in the
case of \hb. Similar considerations apply to $\frac{1}{2}$\ intensity --
the line intensity level at which the virial broadening is computed:
\mgii\ shows modest amplitude normalized shifts (Fig. 
\ref{fig:trendall}).

Eddington ratio estimates are needed to infer the physical meaning of differences  between A and B
spectral types.  We computed the median mass from the medians of the fluxes
of individual sources in each bin and from FWHM measured on the median
composite.  We follow the prescriptions of \citet{assefetal11} and
\citet{shenliu12} to compute \mbh\ from \hb\ and \mgii\ line widths
respectively. Trends are preserved also with older relations linking the
black hole mass to the FWHM and  the continuum at 3100 or 5100 \AA. The
bolometric luminosity needed for \lledd\ has been derived from the
luminosity at 5100 \AA\ following \citet{nemmenbrotherton10}. Values of
bolometric luminosity,  \mbh\ and \lledd\ computed using the BC as a
virial broadening estimator are reported in the first columns Table
\ref{tab:ml}, followed by  \mbh\ and \lledd\ computations using the whole
profile (treating \mgii\ as a single line). \mbh\ values derived from BC
and presented in Table \ref{tab:ml} show a  trend toward decreasing \mbh\
from A1 to A4. A constant median $L$\ in the bins ($\log L \sim 46.0$) 
implies a trend of increasing \lledd\ from A1 to A4. We do not see this
trend if we use \mgii\ measures reflecting  the absence of a clear
correlation between \hb\ and \mgii\ line profile measures in Pop. A. Pop.
B sources show a more regular behavior preserving a trend. This is mainly
due to the definition of B spectral types based on increasing line width.
However, inclusion of VBC leads to an overestimate of $\log$ \mbh \ for
both  \hb\ and \mgii\ derived masses.

Fig. \ref{fig:trendall} shows the behavior of $\delta(\frac{1}{2})$\ and
of the FWHM ratio between the broad component of \mgii\ and \hb\  as a
function of Eddington ratio since the E1 sequence is mainly a sequence of
\lledd.\footnote{Our sample bins have median $L_\mathrm{bol}$  that is
almost constant. This follows from the flux and redshift limits of our
sample, and implies that for any trend as function of \lledd, there is  a
corresponding trend with \mbh. \mbh\ is  a one-way estimator of a quasar
evolutionary status. However, the presence of line shifts is more probably
associated to the relative balance of gravitation and radiation forces so
that we will assume in the following that the relevant physical parameter
is \lledd.}  The parameter  $\delta(\frac{1}{2})$ is
considered  as an appropriate indicator of the dynamical relevance of the
shift because FWHM is a virial broadening estimator of choice. The
$\delta$\ trends at $\frac{1}{4}$\ and $\frac{3}{4}$\ peak intensity as a
function of spectral type are preserved also as a function of \lledd\ and
are not shown again.  The \mgii\ blueshift starts increasing  at $\log$\
\lledd$\sim -0.6$, as the A3 and A4 sources behave differently from all
other sources, as discussed in \citetalias{marzianietal13}. The marginal
blueshift detected at $\frac{3}{4}$\ maximum for spectral type B1 becomes
untraceable at $\frac{1}{2}$\ intensity.  At $\frac{1}{2}$\ maximum  the
\mgii\ profile is mostly unperturbed in all B bins (save B2 that includes
a tiny minority of sources and is probably a confusing mixup), while the
\hb\ profile shows a highly significant shift to the red.

\section{Discussion}
\label{discuss}

\subsection{\mgii\ and \hb\ in the Pop. A and B Context}

Virial estimates of \mbh\ for large quasar samples have been available now
for
more than 10 years \citep[e.g.][]{mclurejarvis02,mcluredunlop04}.
Estimated masses
lie within the range log \mbh $\sim$ 6.0 -- 11.0 with uncertainties i.n
the range 0.2--0.4 dex
at 1$\sigma$\ confidence. The latter uncertainty value appears more
realistic when different
estimates for the  same sources are compared. This is also confirmed  when
we compare
different mean mass estimates for quasar populations (e.g. radio-quiet vs.
radio-loud).
This situation has  led to  claims that FWHM measures contain little or no
information
about black  hole mass \citep{croom11} and others arguing that FWHM
measures for different
lines  are equally valid  \mbh\ estimators \citep{vestergaardetal11}. We
argue that neither
of these claims is valid and that the path to a clearer picture lies
within the concept of a
quasar parameter space like 4DE1. Quasars do not distribute randomly in 4DE1.

How can we move towards more accurate and consistent \mbh\ estimates for
individual quasars and
quasar populations? It is unlikely that inconsistencies are due only to
comparisons involving
spectra of different s/n \citep{assefetal12}. Inconsistencies remain when 
comparing estimates
using the same (i.e. SDSS) spectra. The s/n is certainly an issue but its
effect depends upon
the structure of the line profiles used as virial estimators (see e.g.
\citealt{shenetal08,shenetal11}).
Lines well-fit with single unshifted components are likely the most robust
in the face  of
declining s/n. However many broad lines in type 1 AGN show line shifts and
asymmetries while
others are obviously not well fit with a single component. Fitting lines
with a single Gaussian is
inappropriate because the profile shapes are rarely well approximated by a
single Gaussian.
For example, an attempt to fit the \mgii\ line of spectral type B1 with  a
single Gaussian would
lead to a (bad) fit resulting in an overestimation of FWHM by 700 \kms.
Effects of erroneous assumptions
on fitting function depend on s/n in a way that depends on the intrinsic
shape and on noise properties.
Composites spectra have very high s/n and allow one to focus on systematic
effects with minimal
measurement (statistical) errors. The challenge is to generate composites
in a clear and unbiased way
which means binning physically similar quasars.

If any broad lines serve as virial estimators then \mgii\ and \hb\ are the
safest choices.  Results
presented in this paper suggest that \mgii\ is the  line least affected by
profile shifts and asymmetries
in 90\%\ of all quasars.  We maintain that a first step in the path to
progress requires recognition
that quasars showing FWHM \hb\ $<$ or $>$ 4000\kms\  (Population A and B
respectively) have very
different profile
properties\citep{sulenticetal00b,collinetal06,zamfiretal10}. In other
words, the
dispersion in estimated \mbh\, and resultant \ledd\, measures is not found
by luminosity binning
(at fixed $z$) but rather by binning in the 4DE1 context that reflects
differences in BLR kinematics
and geometry. 

\begin{figure*}
\includegraphics[scale=0.4]{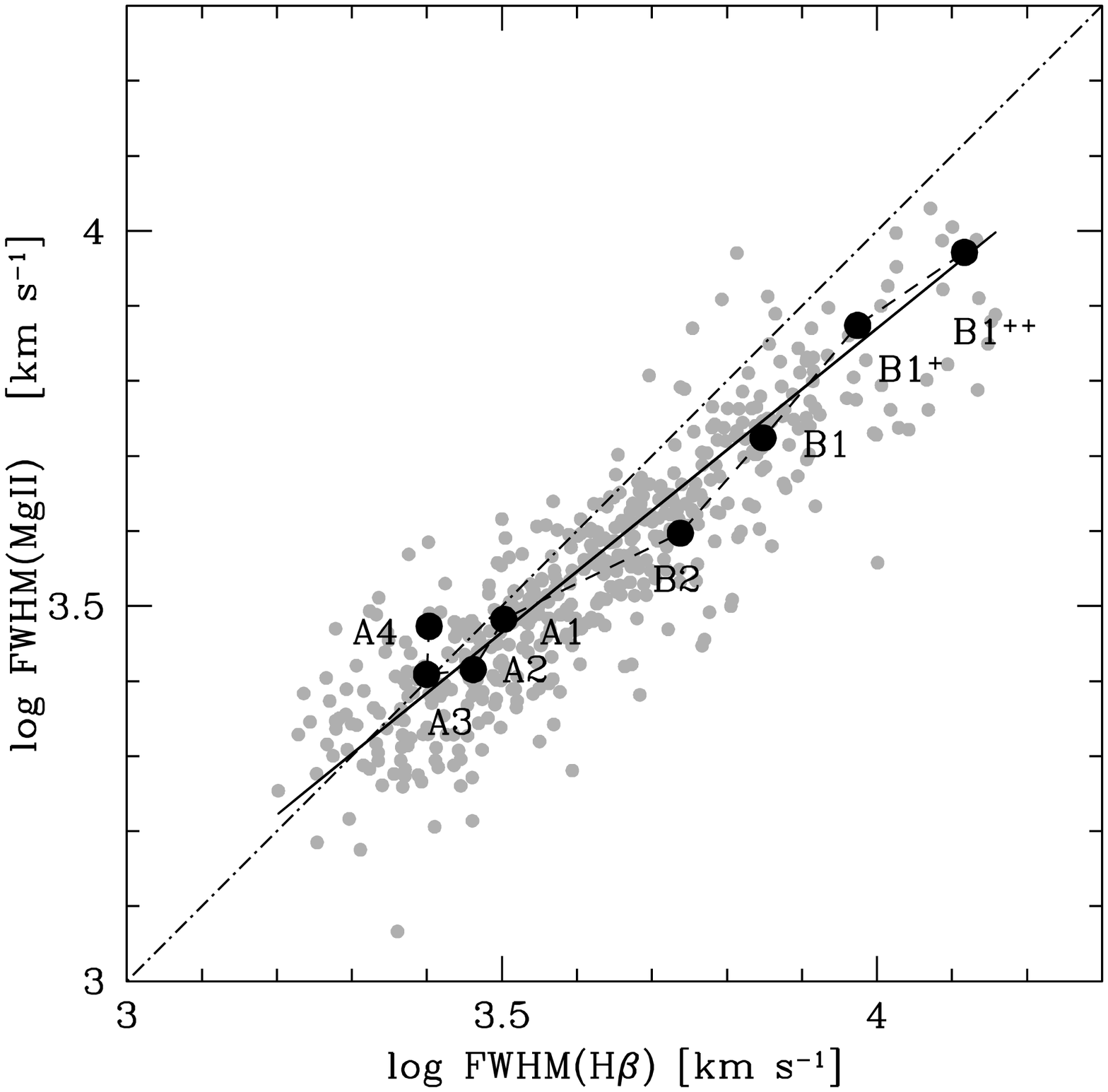}
\includegraphics[scale=0.4]{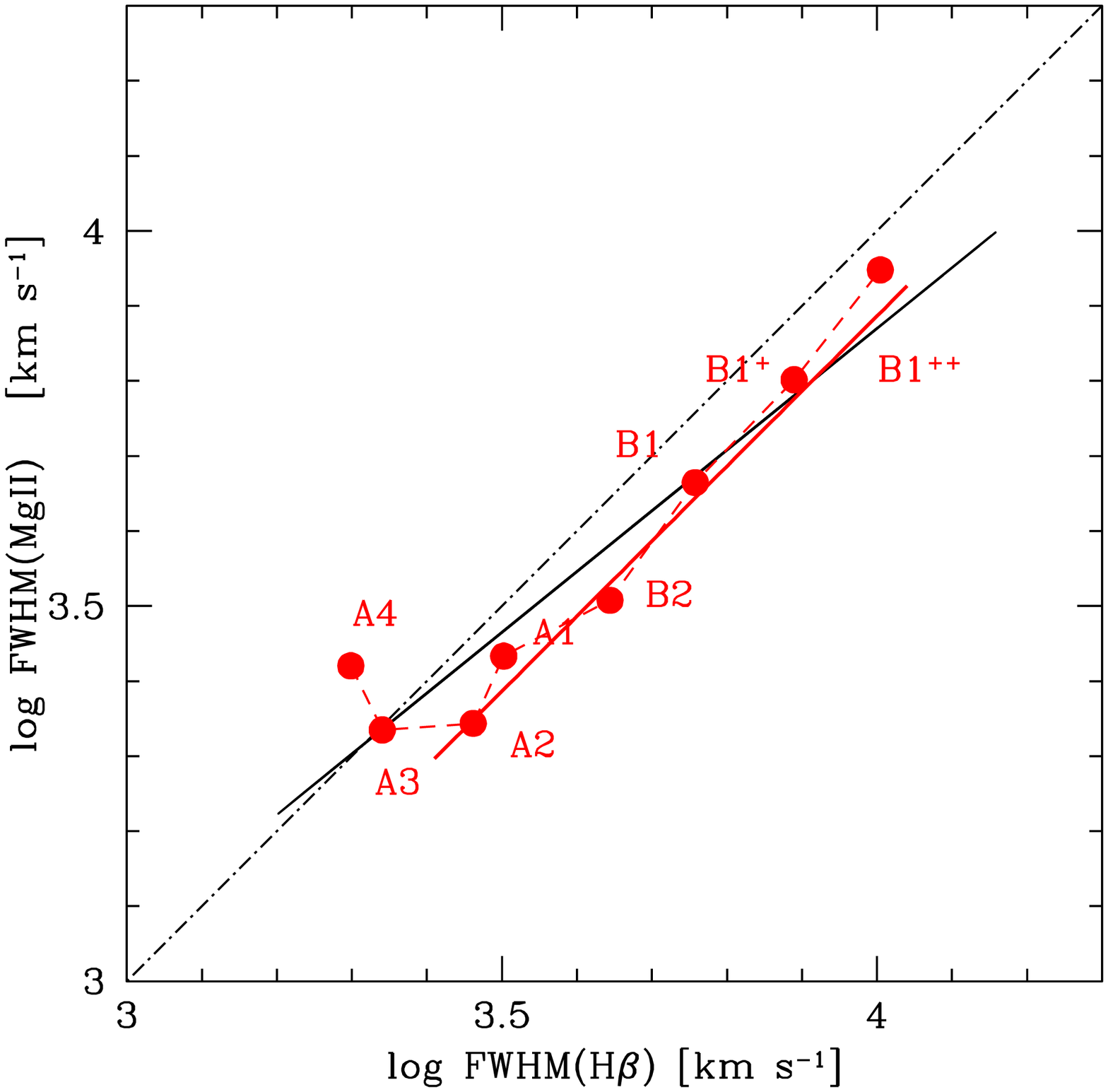}
\caption{FWHM of \mgii\ vs FWHM \hb\ in log scale. Left: full profiles, right: only BC (of single component for \mgii).  Grey dots are the data points of \citet{wangetal09}, with the line representing their best fit. The large black spots are the median values for the full profiles of our median spectra. The dot-dashed lines traces the equality relation, and the thin red line a fixed ratio 0.77. 
\label{fig:w09}}
\end{figure*}

Figure \ref{fig:w09} (left) overlays our composite measures of FWHM \hb\
vs. FWHM \mgii\  (with the \mgii\ doublet treated
as a single line) on the individual source measures from \citet{wangetal09}.
Full profile measures follow  the source distribution and the best-fit
regression of \citet{wangetal09}. The large
scatter at the low FWHM end and the convergence of the regression towards
parity reflect the unusual behavior of sources in bins A3 and A4 where the
FWHM offset between \hb\ and \mgii\ breaks down. The convergence is also
affected at the high
FWHM end where the prominence of the VBC component increases in \hb\ more
than in \mgii.  Figure \ref{fig:w09} (right panel) presents our  BC values
(all Pop. B sources corrected for the VBC) along with the parity line, the
\citet{wangetal09} regression and our best fit to the median composite
values. The main result is that we find no convergence but rather a
constant offset (\mgii\ $\approx$20\% narrower) from bin A2 through all B
bins.  

\subsection{Implications for \mbh\ estimates of quasars: a tentative recipe}
\label{recipe}

The previous analysis suggests the following recipe.

\begin{description}
 \item[1] Identify a source as Population A or B --  {on the basis of
FWHM \hb\ $<$ or $>$ 4000 \kms\ respectively.}

 \item[2a]  Bin  A3 and A4 \mgii\ profiles are  unsuitable as virial
estimators.
  Bin A3 and A4 \hb\ profiles are suitable if corrected for strong \feii\
contamination and for a blue  component.

 \item[2b] A1 and A2. Both \hb\ and \mgii\ can be used as estimators
without any correction.

\item[3] {Pop. B.  Apply correction because of VBC to both \hb\ and
\mgii, as described below.}

\end{description}

A correction factor can be defined as the ratio between observed FWHM and
BC FWHM:

\begin{equation}
\mathrm{FWHM(line)}_\mathrm{vir} = \xi \mathrm{FWHM(line)}_\mathrm{obs}
\end{equation}
where
\begin{equation}
\xi = \frac{\mathrm{FWHM (line)}_\mathrm{BC}}{
\mathrm{FWHM(line)}_\mathrm{obs}} ,
\end{equation}

{\parindent=0pt
and $\mathrm{FWHM(line)}_\mathrm{obs}$\ is measured on the whole profile
(treating the \mgii\ doublet as a single line).  If a single-epoch FWHM
measurement is made without consideration of the intrinsic shape then
$\xi$\ is as reported in Tab. \ref{tab:xi}\  ($\xi$\ values are  derived
from Tab. \ref{tab:specfit} and Tab. \ref{tab:prof}). The $\xi$ factor is
less important for \mgii\ because the VBC component is always weaker. In
the case of \mgii\ a suitable $\xi$ value is $\approx$ 0.85.   Table
\ref{tab:xi} indicates that a simple correction factor $\xi \approx 0.75 - 
0.8$ could be applied to \hb\ once the object is classified as Pop. B, as
suggested by Fig. \ref{fig:w09}.  A simulation degrading the B1 spectrum
with a combination of Gaussian and Poissonian noise shows that  it is
possible to recover a basic estimate of FWHM (with heavy smoothing) down
to s/n $\approx 3$\ \citep[c.f.][]{shenetal08,shenetal11}, even if 
information on profile shape is completely lost. We also note  that the
$\xi$ correction may be accurate  excluding the most luminous sources
where the VBC is   extremely strong i.e., for \lbol $\ga$ 10$^{47}$\ergss\
\citep{marzianietal09}, or the profile is not too irregular as it is the
case of several sources  in B1$^{++}$.  The third column of Tab.
\ref{tab:xi} lists the multiplicative factor needed to convert from
observed FWHM \hb\ to FWHM \mgii\ BC. The last column lists the factor
needed to convert the observed \mgii\ FWHM into FWHM \hbbc. }

The recipe reported above accounts { for our result that the difference
between \mgii\  and \hb\ \mbh\ is positive at large \lledd,} and negative
toward the lower end of the \lledd\ distribution. If no correction is
applied then \mgii\ will be  measured  broader than \hb\ toward  high
\lledd\ values (20\% in A3 and $\approx$ 50\% in bin A4). Therefore {
at high \lledd} the \mgii\ masses will be overestimated with respect to
\hb\ which should actually provide a more reliable estimate. At low
\lledd\ the
effect of the VBC on FWHM \hb\ is stronger yielding an \mbh estimate larger
than the one from \mgii.  The agreement with Fig. 1 of \citet{onkenkollmeier08}  is   quantitative (if rare outlying points in their Fig. 1 are excluded; medians by definition give no or low weight to extremes in a distribution): $\Delta \log$ \mbh = $\log$ \mbh (\mgii) -- $\log$ \mbh (\hb)\    $\approx$ 0.4 dex for bin A4 with $\log$\lledd $\approx$ --0.2, and $\Delta \log$ \mbh $\approx -0.4$\ at   $\log$\lledd $\approx$ --1.7.

Even if A3 and A4 represent only  $\approx$10\%\ of our low redshift
sample, high \lledd\ are discovered  with increasing frequency in
high-$z$, flux limited samples. A recent study  for $z <$1.8 sources 
involving \mgii\ and \ha\ finds a small offset in the  \mbh\ estimates, in
the sense that \mgii-derived masses are slightly larger with respect to
those from \hb\ \citep{matsuokaetal13}. This might indicate an increasing
frequency  of \mgii\ profiles affected by non-virial, outflow motions.

\subsection{Origin of \mgii\ emission and profile blue shifts}
\label{origin}

A major result of this paper is that FWHM  \mgii\ is systematically
narrower than FWHM \hb, excluding types A3 and A4.
This holds for the full profile as well as for {\em all} the components
identified in the lines: i.e. BC, VBC, blueshifted component. The simplest
explanation is that  \mgii\  is emitted at larger radius, on average, than
\hb\ assuming that the
gas is photoionized by a central continuum source.  The  emissivity
$\Sigma$\ of the two lines was computed as a function
of the ionization parameter $U$\ using {\sc cloudy} simulations described
in \S \ref{ratio}. The dependence on $U$ was then converted to radial
distance assuming a constant column density $\log$ \nh = 11 [\cm3]. If we
approximate $\Sigma$ with a power-law  then  $\Sigma \propto r^{-\alpha}$
yields $\Sigma$(\mgii) $\propto r^{-0.1}$ with a steeper trend for \hb,
$\Sigma$(\hb) $\propto r^{-0.4}$, if $\log U \la -0.5$. Elementary
considerations suggest that the different emissivity law can lead to a
different shape in the line wings and hence to a different FWHM for the
two lines. The energy emitted in the line at
distance $r$\ can be written as  $E(r) dr = 4 \pi r^{2} \Sigma(r)
f_\mathrm{c}(r) dr$, where $f_\mathrm{c}$\ is the covering factor. Since
the virial assumption implies that $r \propto \frac{1}{v^{2}}$\ the
specific energy  emitted per unit radial velocity is  $E(v) dv \propto 
v^{-4} v^{2\alpha} f_\mathrm{c}(v) \frac{dv}{v^{3}} \propto  v^{2(\alpha
-q)-3} {dv}$\ if  $ f_\mathrm{c}(r) \propto r^{-2+q}$.  For $\alpha$ =
0.5\ and $q=0$, $E(v) \propto v^{-2}$\ which is the same general power-law
describing  Lorentzian line wings.  A flatter emissivity law will yield a
more peaked profile. Using the emissivity laws for \mgii\ and \hb,
a 20\%\ width ratio is obtained  for reasonable values of $q$\
($-\frac{2}{3} \la q \la \frac{1}{3}$, c.f. \citealt[][]{netzer90}). A
more refined scenario involves the computation of the line profile
following a weak-field approximation  \citep{chenetal89,sulenticetal98}.
Using the {\sc cloudy}-computed emissivity laws for \hb\ and \mgii, the
profile of \mgii\ turns out to be systematically narrower than  \hb\ by
$\ga$10\%.  The width ratio can be $\approx$20\%\ if the inner radius of
the emitting disk is $\approx$ 250 gravitational radii. Therefore,  the
line width ratio and the Lorentzian-like  profile shape
can be explained in the context of virialialized gas motions.

The \mgii\ blueshift is most straightforwardly attributed to outflow of
the line emitting gas with preferential
obscuration of the receding part of the flow \citetalias{marzianietal13}. 
A blueshift is suspected in the B1 composite (Fig. \ref{fig:hbmgb}) but is
not significant at a 2$\sigma$ confidence level. The small amplitude of
the shifts does not rule out a change in the intrinsic multiplet ratio as
an alternative explanation in Pop. B sources: a value $\mathbb{R} = 2$
would yield an effective  wavelength close to the one of the
$^{2}P_{\frac{3}{2}} \rightarrow ^{2}S_{\frac{1}{2}}$ transition, 2796.35.
If the effective wavelength is close to this value,  then any systematic
blueshift $\ltsim 300$\kms\ will become insignificant. If the blueshift is
associated with a low-density high-ionization component then a value as
large as $\mathbb{R} \approx$1.7 could be possible. However, the absence
of any blueshift in bin A1 that is the simplest profile to analyze (1
component, weak \feii) argues against this suggestion.

Outflows driven by line or ionizing photon pressure  can accelerate the line-emitting gas to a terminal velocity that is inversely proportional to the square root of the column density \nc\ and directly proportional to the Eddington ratio i.e., 
\begin{equation}
\label{eq:vtr}
v_\mathrm{t} \propto \left(\frac{1}{N_\mathrm{c}}   \frac{ L}{L_\mathrm{Edd}}\right)^{\frac{1}{2}} v_\mathrm{Kepl}
\end{equation}
where $v_\mathrm{Kepl}$\ is the Keplerian velocity at the outflow starting radius.  

If real the modest shift observed in the median spectra of B1
and B1$^{++}$\ could be interpreted as the terminal velocity of
a radiation driven outflow. This shift could  occur even if
\lledd $\sim 0.01$ in gas with moderate column density \nc $\sim
10^{23}$\cmq\ (provided that the gas remains optically thick to the ionising continuum) that   may in turn support  the idea that resonant lines are driving, at least in part, the acceleration of the outflow \citep{progaetal00,gangulyetal07,sulenticetal07}. There is an intriguing similarity in the \mgii\ and \civ\ behavior that may be related to the resonant nature of both lines but the effects on \mgii\ appear by far less relevant than on \civ\ since the median \mgii\ profile retains an overall symmetric appearance. The B1 and  B1$^{++}$\ profiles are {\em on average} only slightly perturbed by blueshifted emission: low-\nc\ gas might  produce some scatter
among B sources since it is most strongly affected by radiative forces
and may become unbound following a continuum luminosity increase
\citep{netzermarziani10}. {  B1$^{++}$\ sources are likely to be some
of the lowest \lledd\ sources. They are unstable in continuum emission and 
often show strongly variable emission line profiles with
irregular shapes \citep{lewisetal10}.} { They are also 
rare ($\la 3$\%\ in the present sample) and  are interpreted as the most
evolved, perhaps dying or starving quasars in the 4DE1 sequence. }  An
example involves the well-studied FRII source \object{3C 390.3} which
belongs to bin B1$^{++}$ with line profiles characterized by a prominent
blueshifted peak that produces a significant net profile blueshift.
Another example with extreme velocity offset and large FWHM involves 
\object{SDSS 0956+5128} \citep{steinhardtetal12}. Even if the
interpretation of 3C 390.3 is currently a subject of debate
\citep[e.g.][]{dietrichetal12,zhang11}, an application of the method of
\citet{negreteetal12} indicates relatively low-density and high-ionization
consistent with the overall properties of the blueshifted component
identified in the spectra of many quasars \citep{marzianietal10}. { A
significant fraction of B1$^{++}$\ sources in the present sample indeed
show \mgii\ profiles with large blueshifts/blueward asymmetries. }
Therefore, the marginal blueshifts of spectral type B1$^{++}$ \ median
composite may reflect the somewhat erratic nature of low-density
outflowing gas detected { in sources
whose unstable accretion rate may induce significant continuum
fluctuations. }

The high s/n of our composite spectra reveals a semi-broad component in
the \oiiiopt\ profiles and on the blue side of \hbnc. B1 offers the most
striking case along with \object{PG 1201+436}. In the B1 \hb\ composite
spectrum the excess flux  is explained as a semi-broad component with 
width and shift similar to the \oiiiopt\ semi-broad component. That
feature does not belong to the broad profile and would induce a blueshift
of $\sim -100$ \kms\ into the BC model or into the measures of the broad
line profile if not adequately taken into account. A similar
interpretation is also possible for the small blueshifted ``bump''
observed in the \hb\ and \mgii\ profiles of PG 1201+436 (blue line in Fig.
\ref{fig:pg1201}). The bump is modeled with a Gaussian component of width
1500 -- 2500 \kms\ and shift $-1500 - -2000$ \kms. The inferred velocities
and widths appear rather extreme to other cases where a semi-broad
component of \oiiiopt\ is isolated. We can speculate that also this source
is showing outflow largely in a region that might be at the boundary
between the BLR and the narrow-line region \citep{zamanovetal02}.  A more
refined analysis of \object{PG 1201+436}  is deferred to later work
(Sulentic et al., in preparation).

\subsection{Radio quiet and radio loud}
\label{rl}


Core-dominated  and lobe-dominated sources show a fairly symmetric profile, with  evidence of a redshifted base due to the VBC that is especially prominent in CD sources \citep[e.g.,][]{punslyzhang11}.  The red excess is clearly detected also in \mgii\ \citep[a very convincing case is provided by the blazar 3C 279, ][]{punsly13} although the A.I.\ is significantly lower than \hb. The absence of any hint of blueshifts in RL profiles could be related to the full suppression of an accretion disk wind, or it could be   due to orientation. B1 bin  was defined on a limited range of line width.  Therefore we cannot expect to see the systematic FWHM difference mentioned above. In addition the fraction of FRII increases toward bin B1$^{++}$\, i.e., broader sources. In bin B1 we are probably considering a subsample of FRII that show narrower profile then the majority of the population. If the FWHM is affected by orientation as expected \citep[e.g.,][]{willsbrowne86,sulenticetal03,rokakietal03,runnoeetal12,zamfiretal08}, we are observing  sources that tend to be seen at smaller viewing angle than the general FRII population. If the wind flow follows a bent path over the accretion disk \citep{elvis00} or if the flow is constrained closer to the disk plane, the absence of a blueshift might be therefore related more to  orientation than to  full suppression of disk winds in RL sources by the pressure of a cocoon associated with the relativistic ejections \citep{normanmiley84}. Disk winds are probably not fully suppressed in radio-loud sources, as indicated by sources like 3C390.3 and by the discovery of fast outflows in radio-loud AGNs \citep{tombesietal10}.

\subsection{\mgii\ virial broadening  and virial product}
\label{virial}

The FWHM difference between \hb\ and \mgii\ is very important. From the
data presented in this paper, with the exception of spectral types A3 and
A4, the FWHM(\hb) $\gtsim$ FWHM (\mgii) both if the whole profile is
considered or if only the broad component is isolated.  The FWHM of a line
by itself is of limited usefulness in computing \mbh\ following Eq.
\ref{eq:vir}. A virial broadening estimator (let it be a line FWHM or
velocity dispersion) should be associated to a typical distance to put in
use the simple virial relation that is customarily employed in \mbh\
calculations
\citep{mcluredunlop04,vestergaardpeterson06,shenetal08,trakhtenbrotnetzer12}.
In principle, there are two main possibilities in the interpretation of
the width differences (not mutually exclusive).  The smaller  \mgii\ FWHM
may imply that (1)   \mgii\ is emitted farther out from the central
continuum source according to the virial law $\Delta v \propto r^{-1/2}$.
If the FWHM ratio is $\approx$ 0.8, then \mgii\ should occur at a distance
significantly larger than \hb, by a factor 1.4; (2)  only part of the gas
emitting \hb\ is emitting \mgii.  If (1) applies  the scaling relations
involving \mgii\ should be defined with some care since \rb\ defined from
reverberation mapping of \hb\ would suggest smaller distances.  In the
second case, \mgii\ can be interpreted as perhaps the ``best'' estimator,
associated to gas very optically thick that responds to continuum changes.
Reverberation mapping studies of \mgii\ are  not yet conclusive, and 
they leave open the possibility that the average emitting distance of
\mgii\ is larger than for \hb\ \citep{metzrothetal06,treveseetal07,woo08}.
In addition to the theoretical considerations of \S \ref{origin}\ that
favor a larger distance for \mgii\ emission, there is  empirical evidence
provided by the difference between the \mgii\ and \hb\ profile shape.  For
Pop. B sources, the \hb\ VBC gives rise to a stronger redward asymmetry
than in \mgii, and suggests that \mgii\ emission is occurring farther away
from the central continuum source. This is consistent with the observation
of smaller FWHM(\mgii) for both BC and whole profile  for spectral type A1
and A2 that do not show appreciable VBC emission.
The implication of a larger emissivity-weighted distance for \mgii\ is a
shift in the correlation between \rb\ (computed from \hb) and the
luminosity at 3000 \AA, by  $\approx$ 0.16 dex in \rb\ in Eq. 7 and Fig. 4
of \citet{trakhtenbrotnetzer12}. However, Eq. 12 of 
\citet{trakhtenbrotnetzer12} will be unaffected   since the virial factor
is scaled to yield the \mbh\ derived from \hb.  The difference is that
$\cal{F}$ should be $\approx 1.0$\ and not $\cal{F}  \approx$1.3 - 1.4
that follows from the assumption that \hb\ and \mgii\ are emitted at the
same distance.
Scalings adopted in all previous studies assumed that a single $\cal{F}$ \
value is appropriate for all quasars. This assumption is most likely
incorrect, especially if non-gravitational forces are at play. Previous
work suggested a significant $\cal{F}$\ difference between Pop. A and B
\citep{collinetal06}. A much needed improvement is therefore an evaluation
of $\cal{F}$  as a function of spectral type or \lledd. This is however
beyond the scope of the present paper.

\section{Conclusion}

Both  \hb\ and \mgii\ appear to be suitable for \mbh\ virial  estimates  of black hole mass at least for a large fraction of quasars. Both low ionization lines, however, show profile shifts and asymmetries that must be taken into
account if our goal is to improve the poor present-day accuracy of most
estimates. { Right now, this analysis suggests that masses for Pop. B sources
based on \hb\ are likely systematically overestimated. } Further
progress therefore requires subdivision of large quasar samples into Pop.
A and B with further subdivision into spectral types facilitating
consideration of \mbh\ and/or \lledd\ trends. 
4D Eigenvector 1 currently offers the most effective context for  binning. Such source discrimination cannot be
accomplished with $L$, $z$\ binning because sources with very different
line profile properties show similar \lbol\ even when they host
super-massive black holes  with significantly different masses (i.e. the
highest accretors are connected  with Pop. A sources involving the smallest black hole
masses--effectively blurring any expected mass-luminosity correlation for
quasars).


Our ability to distinguish
between Pop. A and B sources is fortunately preserved even
at low s/n. We defined simple prescriptions  with the goal
of providing a better approximation of the virial
broadening for single-epoch observations.

\begin{acknowledgements}
PM acknowledges Junta de Andaluc\'{\i}a, through
grant TIC-114 and the Excellence Project P08-TIC-3531, and  the Spanish Ministry for Science and Innovation through grants AYA2010-15169 for supporting a sabbatical stay at IAA-CSIC. I.~P. - F. acknowledges the postdoctoral fellowship grants 145727
and 170304 from CONACyT Mexico. Funding for the SDSS and SDSS-II has been provided by the Alfred P. Sloan Foundation, the Participating Institutions, the National Science Foundation, the U.S. Department of Energy, the National Aeronautics and Space Administration, the Japanese Monbukagakusho, the Max Planck Society, and the Higher Education Funding Council for England. The SDSS Web Site is http://www.sdss.org/. The SDSS is managed by the Astrophysical Research Consortium for the Participating Institutions. The Participating Institutions are the American Museum of Natural History, Astrophysical Institute Potsdam, University of Basel, University of Cambridge, Case Western Reserve University, University of Chicago, Drexel University, Fermilab, the Institute for Advanced Study, the Japan Participation Group, Johns Hopkins University, the Joint Institute for Nuclear Astrophysics, the Kavli Institute for Particle Astrophysics and Cosmology, the Korean Scientist Group, the Chinese Academy of Sciences (LAMOST), Los Alamos National Laboratory, the Max-Planck-Institute for Astronomy (MPIA), the Max-Planck-Institute for Astrophysics (MPA), New Mexico State University, Ohio State University, University of Pittsburgh, University of Portsmouth, Princeton University, the United States Naval Observatory, and the University of Washington.

\end{acknowledgements}

\clearpage
\bibpunct{(}{)}{;}{a}{}{,} 
\bibstyle{aa} 

\clearpage

\begin{table*}
\setlength{\tabcolsep}{1pt}
\begin{center}
\caption{Sample: number of sources in each spectral type\label{tab:numbers}}
\begin{tabular}{ccccccccl}\hline\hline
\multicolumn{1}{l}{Sp. Type} & \multicolumn{1}{c}{$N_\mathrm{tot}$}  &\multicolumn{1}{c}{$N_\mathrm{RQ}^\mathrm{a}$}  & \multicolumn{1}{c}{$N_\mathrm{FIRST}^\mathrm{b}$} &\multicolumn{1}{c}{$N_\mathrm{FRII}^\mathrm{c}$} & 
\multicolumn{1}{c}{N$_\mathrm{CD}^\mathrm{d}$} \\ \hline
A1	&	{\bf 97}	&	{\bf 79}	&	24	&	10	&	8	\\
A2	&	{\bf 156}	&	{\bf152}	&	20	&	1	&	3	\\
A3	&	{\bf 43}	&	{\bf 39}	&	14	&	0	&	4	\\
A4	&	{\bf 15}	&	{\bf 13}	&	4	&	0	&	2	\\
B1	&	{\bf 218}	&	{\bf 179}	&	58	&	{\bf 23}	&	{\bf 16}	\\
B1$^{+}$\	&	{\bf 115}	&	{\bf 95}	&	27	&	11	&	9	\\
B1$^{++}$\	&	{\bf 17}	&	9	&	11	&	4	&	4	\\
B2	&	{\bf 19}	&	{\bf 16}	&	5	&	0	&	3	\\
All	&	680	&	582	&	163	&	49	&	49	\\ \hline
\end{tabular}
\end{center}
\begin{list}{}{}
\item[$^\mathrm{a}$]{Number of radio-quiet sources defined as $N_\mathrm{RQ}$= $N_\mathrm{tot}$ - $N_\mathrm{CD}$ - $N_\mathrm{FRII}$. }
\item[$^\mathrm{b}$]{{\bf Number of FIRST-detected sources. Note that $N_\mathrm{FIRST} >  N_\mathrm{CD}  +  N_\mathrm{FRII}$.}}
\item[$^\mathrm{c}$]{{ Number of FIRST-detected sources with compact morphology and emitted rest frame power $\log P_{\nu} \ge 31.6$ ($P_{\nu}$ in units of erg s$^{-1}$ Hz$^{-1}$). }}
\item[$^\mathrm{d}$]{{ Number of FIRST-detected sources with extended morphology. }}
\end{list}
\end{table*}

\tiny

\begin{landscape}
\begin{table*}\setlength{\tabcolsep}{3pt}
\begin{center}
\caption{Measured Parameters on the \hb\ and \mgii\ Profiles of Composite Spectra \label{tab:prof}}
\begin{tabular}{lccccccccccccccccccl}\hline
\multicolumn{1}{l}{Sp. Type} &  \multicolumn{6}{c}{\hb} && \multicolumn{6}{c}{\mgii}\\ \cline{2-8} \cline{10-16}
& \multicolumn{1}{c}{FWHM$^\mathrm{a}$} &\multicolumn{1}{c}{A.I.$^\mathrm{b}$}& \multicolumn{1}{c}{$\kappa^\mathrm{c}$} &  
\multicolumn{1}{c}{c($\frac{1}{4}$)$^\mathrm{a,d}$} & \multicolumn{1}{c}{c($\frac{1}{2}$)$^\mathrm{a,d}$} & \multicolumn{1}{c}{c($\frac{3}{4}$)$^\mathrm{a,d}$} & \multicolumn{1}{c}{c(0.9)$^\mathrm{a,d}$}  
&& \multicolumn{1}{c}{FWHM$^\mathrm{a}$} &\multicolumn{1}{c}{A.I.$^\mathrm{b}$} &    \multicolumn{1}{c}{$\kappa^\mathrm{c}$}&
\multicolumn{1}{c}{c($\frac{1}{4}$)$^\mathrm{a,d}$} & \multicolumn{1}{c}{c($\frac{1}{2}$)$^\mathrm{a,d}$} & \multicolumn{1}{c}{c($\frac{3}{4}$)$^\mathrm{a,d}$} & \multicolumn{1}{c}{c(0.9)$^\mathrm{a,d}$} 
\\ \hline 
\multicolumn{16}{c}{{\em Pop. A Full Sample}} \\
A1	&	3180$\pm$90	&	0.00$\pm$0.04	&	0.33$\pm$	0.02	&	50$\pm$110	&	50$\pm$60	&	50$\pm$40	&	110$\pm$50	&&	3040$\pm$80	&	0.01$\pm$0.04	&	0.36$\pm$0.02	&	80$\pm$100	&	80$\pm$50	&	70$\pm$40	&	60$\pm$50	\\
A2	&	2900$\pm$210	&	0.00$\pm$0.10	&	0.33$\pm$	0.04	&	-20$\pm$240	&	-20$\pm$110	&	-20$\pm$90	&	-40$\pm$110	&&	2600$\pm$160	&	0.01$\pm$0.09	&	0.38$\pm$0.04	&	-60$\pm$200	&	-70$\pm$90	&	-70$\pm$80	&	-80$\pm$100	\\
A3	&	2510$\pm$200	&	-0.15	$\pm$0.10	&	0.30$\pm$	0.04	&	-330$\pm$230	&	-90$\pm$110	&	-10$\pm$80	&	30$\pm$90	&&	2560$\pm$160	&	0.01$\pm$0.09	&	0.38$\pm$0.04	&	-140$\pm$200	&	-140$\pm$90	&	-150$\pm$80	&	-160$\pm$100	\\
A4	&	2530$\pm$250	&	-0.23$\pm$0.10	&	0.27$\pm$	0.04	&	-530$\pm$250	&	-150$\pm$130	&	10$\pm$80	&	80$\pm$100	&&	2970$\pm$210	&	0.01$\pm$0.09	&	0.35$\pm$0.04	&	-240$\pm$220	&	-250$\pm$120	&	-250$\pm$90	&	-260$\pm$130	\\
\multicolumn{16}{c}{{\em Pop. B Full Sample}} \\
B1	&	7050$\pm$180	&	0.09$\pm$0.04	&	0.38$\pm$0.02	&	660$\pm$230	&	220$\pm$100	&	140$\pm$80	&	240$\pm$110	&&	5300$\pm$120	&	0.07$\pm$0.03	&	0.42$\pm$0.02	&	230$\pm$140	&	10$\pm$70	&	-40$\pm$70	&	-60$\pm$90	\\
B1$^+$	&	9420$\pm$220	&	0.08$\pm$0.03	&	0.40$\pm$0.02	&	790$\pm$220	&	380$\pm$120	&	250$\pm$110	&	430$\pm$150	&&	7480$\pm$170	&	0.06$\pm$0.03	&	0.41$\pm$0.02	&	380$\pm$180	&	140$\pm$100	&	70$\pm$90	&	50$\pm$120	\\
B1$^{++}$	&	13080$\pm$720	&	0.04$\pm$0.06	&	0.43$\pm$0.04	&	600$\pm$570	&	400$\pm$360	&	290$\pm$350	&	480$\pm$480	&&	9360$\pm$490	&	0.02$\pm$0.06	&	0.45$\pm$0.04	&	100$\pm$360	&	30$\pm$240	&	0$\pm$250	&	-10$\pm$350	\\
B2	&	5470$\pm$340	&	0.15$\pm$0.14	&	0.35$\pm$0.06	&	680$\pm$650	&	60$\pm$180	&	-30$\pm$160	&	-100$\pm$200	&&	3950$\pm$230	&	0.10$\pm$0.11	&	0.39$\pm$0.05	&	120$\pm$360	&	-130$\pm$130	&	-180$\pm$120	&	-200$\pm$150	\\
\multicolumn{16}{c}{{\em Radio Loud}} \\
B1 CD	&	7060$\pm$460	&	0.16$\pm$0.04	&	0.36$\pm$0.02	&	1230$\pm$230	&	440$\pm$230	&	310$\pm$80	&	500$\pm$240	&&	5690$\pm$330	&	0.10$\pm$0.03	&	0.39$\pm$0.02	&	590$\pm$160	&	270$\pm$170	&	180$\pm$70	&	160$\pm$210	\\
B1 FRII	&	6790$\pm$410	&	0.09$\pm$0.05	&	0.37$\pm$0.02	&	850$\pm$260	&	420$\pm$200	&	330$\pm$80	&	620$\pm$250	&&	5500$\pm$300	&	0.05$\pm$0.03	&	0.43$\pm$0.02	&	230$\pm$120	&	100$\pm$160	&	50$\pm$80	&	30$\pm$210	\\
\hline
\end{tabular}
\begin{list}{}{}
\item[$^\mathrm{a}$]{In units of \kms.}
\item[$^\mathrm{b}$]{{ Dimensionless asymmetry index computed at $\frac{1}{4}$ peak intensity.}}
\item[$^\mathrm{a}$]{{ Dimensionless kurtosis index defined as the ratio between the full width at $\frac{3}{4}$\ and $\frac{1}{4}$\ maximum.}}
\item[$^\mathrm{d}$]{{ Centroids at  $\frac{1}{4}$,  $\frac{1}{2}$,  $\frac{3}{4}$\ and $\frac{9}{10}$ maximum intensity.}}
\end{list}
\end{center}
\end{table*}
\end{landscape}

\begin{landscape}
\begin{table*}
\begin{center}
\caption{Derived Quantities from the \hb\ and \mgii\ Profiles  Composite Spectra Multicomponent Analysis \label{tab:specfit}}
\setlength{\tabcolsep}{1pt}
\begin{tabular}{lccccccccccccccc}\hline
\multicolumn{1}{c}{Sp. Type} &  \multicolumn{6}{c}{\hb} && \multicolumn{6}{c}{\mgii}\\ \cline{2-7} \cline{9-14}
 &\multicolumn{1}{c}{BC$^\mathrm{a}$}& \multicolumn{1}{c}{Shift$^\mathrm{b}$} &  
\multicolumn{1}{c}{FWHM$^\mathrm{b}$} &\multicolumn{1}{c}{VBC$^\mathrm{a}$}& \multicolumn{1}{c}{Shift$^\mathrm{b}$} &  
\multicolumn{1}{c}{FWHM$^\mathrm{b}$} &  &\multicolumn{1}{c}{BC$^\mathrm{a}$} & 
\multicolumn{1}{c}{Shift$^\mathrm{b}$} & \multicolumn{1}{c}{FWHM$^\mathrm{b}$}  &\multicolumn{1}{c}{VBC$^\mathrm{a}$} &\multicolumn{1}{c}{Shift$^\mathrm{b}$} &  
\multicolumn{1}{c}{FWHM$^\mathrm{b}$}\\
 \hline\hline
\multicolumn{14}{c}{{\em Pop. B Full Sample}} \\
B1	           &	46		&	20$^\mathrm{c}$	&	5720$\pm 100$	&	51	&	1740$\pm 150$	&	15600$\pm 150$	&&	63	&	-165$\pm 50$	&	4620$\pm 100$	 	&	36	&	2160$^\mathrm{c}$&	11500$\pm 300$	\\
B1$^{+}$\	   &	46		&	5$^\mathrm{c}$	&	7750$\pm 200$	&	45	&	1920$\pm 190$&	15700$\pm 260$	&&	68	&	-60$\pm 80$	&	6330$\pm 100$	 	&	51	&	1520$^\mathrm{c}$&	14500$\pm 1230$	\\
B1$^{++}$\  &	28		&	-165$^\mathrm{c}$	&    10100$\pm 420$&	65	&	915$\pm 70$	&	15830$\pm 860$	&&	82	&	-105$^\mathrm{c}$	&	8870$\pm 420$ 	&	20	&1305$^\mathrm{c}$	&	13000$^\mathrm{c}$	\\

B2	&	33		&	-120$\pm 60$	&	4410$\pm 100$	&	40	&	1950$\pm 280$	&	13400$\pm 600$	&&	43	&	-240$\pm 40$	&	3220$\pm 240$ 	&	38	&	1420$\pm 540$	&	10000$^\mathrm{c}$	\\				
\\
PG 1201+436 &    42$^\mathrm{d}$ & -5$^\mathrm{c}$ & 7500$^\mathrm{c}$ & 179$^\mathrm{d}$ & 2125$\pm $370 & 11800$\pm$770 && 118$^\mathrm{d}$ & 0$^\mathrm{c}$ & 7220$\pm 590$& 94$^\mathrm{d}$ & 1790$\pm$330 & 9780$\pm$740\\ 
\multicolumn{14}{c}{{\em Radio Loud}} \\
B1 CD	&	34		&	80$^\mathrm{c}$	&	5330$\pm 200$	&	50	&	1935$\pm 190$ 	&	13160$\pm 260$	&&	60	&	45$\pm 60$	&	4680$\pm 120$	&	51	&	1630$^\mathrm{c}$	&	11000$^\mathrm{c}$	\\
B1 FRII	&	47		&	235$\pm$55	&5370$\pm 130$&	60	&	1645$\pm 170$	&	15450$\pm 530$	       &&	74	&	-55$\pm 70$	&	4790$\pm 200$	&	37	&	1630$^\mathrm{c}$	&	11000$^\mathrm{c}$	\\ \hline
\end{tabular}                   							
\begin{list}{}{}
\item[$^\mathrm{a}$]{~Line intensity normalized to the continuum at 5080 \AA. 
The value  roughly corresponds to the rest-frame equivalent width in \AA\ for \hb.  
{\bf Intensity ratios within each spectral type can be computed from the reported values. }}
\item[$^\mathrm{b}$]{In units of \kms.}
\item[$^\mathrm{c}$]{~Kept fixed or allowed to vary within a narrow interval in {\sc specfit} analysis.}
\item[$^\mathrm{d}$]{~In units of 10$^{-15}$ erg s$^{-1}$ cm$^{-2}$ \AA$^{-1}$ }

\end{list}
\end{center}
\end{table*}
\end{landscape}

\hoffset=-1cm
\begin{table*}
\caption{Median bolometric luminosity, black hole mass and Eddington ratio \label{tab:ml}}
\setlength{\tabcolsep}{3pt}
\begin{center}
\begin{tabular}{lccccccccccccc}\hline
\multicolumn{1}{c}{Sp. Type} &&  \multicolumn{4}{c}{BC only} && \multicolumn{4}{c}{Whole profile}\\ \cline{4-7} \cline{9-12} 
& \multicolumn{1}{c}{$L_\mathrm{bol}^\mathrm{a}$} &&\multicolumn{1}{c}{\mbh$^\mathrm{b}$}& \multicolumn{1}{c}{\mbh$^\mathrm{b}$} &  
\multicolumn{1}{c}{\lledd$^\mathrm{c}$} & \multicolumn{1}{c}{\lledd$^\mathrm{c}$} 
&&\multicolumn{1}{c}{\mbh$^\mathrm{b}$}& \multicolumn{1}{c}{\mbh$^\mathrm{b}$} &  
\multicolumn{1}{c}{\lledd$^\mathrm{c}$} & \multicolumn{1}{c}{\lledd$^\mathrm{c}$} \\
& & & \hb\ & \mgii & \hb\ & \mgii &&  \hb\ & \mgii & \hb\ & \mgii  \\ \hline\hline
\multicolumn{12}{c}{{\em Pop. A Full Sample}} \\
A1	&46.18	&&8.67	&8.62	&-0.60 	&-0.56 &&8.67	&8.71&	-0.60 	&-0.64\\
A2	&46.16	&&8.57	&8.45	&-0.53 	&-0.40 &&8.57	&8.57	&-0.52	&-0.52\\
A3	&46.15	&&8.33	&8.41	&-0.28	 &-0.37 &&8.45	&8.53	&-0.41	&-0.49\\
A4	&46.20	&&8.27	&8.56	&-0.18	&-0.47&& 8.48 &8.66	&-0.39	&-0.57\\
\multicolumn{12}{c}{{\em Pop. B Full Sample}} \\
B1	&46.13	&&9.15	&8.98	&-1.13	&-0.96 &&9.33	&9.08	&-1.31&	-1.06\\
B1$^{+}$	&45.96	&&9.29	&9.06	&-1.44	&-1.21 &&9.46	&9.18	&-1.61&	-1.33\\
B1$^{++}$&45.96	&&9.53	&9.14	&-1.68	&-1.29 &&9.75	&9.35	&-1.90&	-1.50\\
B2	&46.23	&&8.99	&8.74	&-0.87	&-0.62 &&9.17	&8.89	&-1.05&	-0.77\\
\hline\end{tabular}
\begin{list}{}{}
\item[$^\mathrm{a}$]{Decimal logarithm of bolometric luminosity in units of \ergss. A bolometric correction computed following \citet{nemmenbrotherton10} was applied to the 5100 \AA\ specific luminosity.}
\item[$^\mathrm{b}$]{Decimal logarithm of black hole mass in solar units. \mbh\ has been estimated from the relation  of \citet{shenliu12} for \mgii\ and from the relation \citet{assefetal11} for \hb.}
\item[$^\mathrm{c}$]{Decimal logarithm of Eddington ratio.}
\end{list}
\end{center}
\end{table*}

\hoffset=-1cm
\begin{table*}
\caption{Conversion factors \label{tab:xi}}
\setlength{\tabcolsep}{3pt}
\begin{center}
\begin{tabular}{lcccc}\hline\hline
  & $\xi$(\hb)$^\mathrm{a}$  &         $\xi$(\mgii)$^\mathrm{a}$      &      \mgiibc/     &   \mgii\ obs/    \\
  &  & & \hb\ obs &  \hbbc\       \\                \hline
A2         & 1.00  &  1.00  & 0.76 &   0.90   \\
A1         & 1.00  &  1.00  & 0.85 &   0.96   \\
B1         & 0.81  &  0.87 &  0.66 &   0.93      \\ 
B1 CD   & 0.75  &  0.82 &  0.66  &   1.07    \\   
B1 FRII & 0.79  &  0.87 &  0.71  &   1.02       \\ 
B1$^{+}$     & 0.82 &   0.85 &  0.67  &   0.97      \\ 
B1$^{++}$    & 0.77  &   0.95 & 0.68  &  0.93 \\
\hline\end{tabular}
\begin{list}{}{}
\item[$^\mathrm{a}$]{{Conversion factor $\xi$\ is the FWHM of BC divided by the FWHM of the full (``observed'') profile. The \mgii\ doublet is treated as a single feature.}}
 \end{list}

\end{center}
\end{table*}

\pagebreak\eject
\newpage\clearpage

\vfill\break
\pagebreak
\newpage

\hoffset=0.5cm

\vfill\break
\pagebreak
\newpage


\vfill\break
\pagebreak
\newpage

\clearpage

\end{document}